\documentclass{jfm}
\usepackage{soul}
\usepackage{graphicx}
\usepackage{bm}
\usepackage{newtxtext}
\usepackage{newtxmath}
\usepackage{natbib}
\usepackage{hyperref}
\hypersetup{
	colorlinks = true,
	urlcolor   = blue,
	citecolor  = black,
}

\newcommand{\RomanNumeralCaps}[1]
\linenumbers

\title{Symmetry-breaking bifurcations and sub-harmonic lock-in of a flexible splitter plate in cylinder wake flow}

\author{Baiyang Song\aff{1},
	Huan Ping\aff{2},
    Wenli Chen\aff{3},
	Yong Cao\aff{1,4} \thanks{\email{yongcao@sjtu.edu.cn}}
		\and  Dai Zhou\aff{1,5}
        \corresp{\email{zhoudai@sjtu.edu.cn}}
    }

\affiliation{\aff{1}State Key Laboratory of Ocean Engineering, School of Ocean and Civil Engineering, Shanghai Jiao Tong University, Shanghai, 200240, China
	\aff{2}College of Ocean Science and Engineering, Shanghai Maritime University, Shanghai, 201306, China
    \aff{3}School of Civil Engineering, Harbin Institute of Technology, Harbin, 150001, China
    \aff{4}Chongqing Research Institute, Shanghai Jiao Tong University, Chongqing, 401135, China
    \aff{5}Shenzhen Research Institute of Shanghai Jiao Tong University, Shenzhen 518063, China}

\begin{document}
	\maketitle
	
	\begin{abstract}

		This paper investigates the flow past a flexible splitter plate attached to the rear of a fixed circular cylinder at a low Reynolds number of 150. 
         A systematic exploration of the plate length ($L/D$), flexibility coefficient ($S^{*}$), and mass ratio ($m^{*}$) reveals new laws and phenomena.
        The large-amplitude vibration of the structure is attributed to a resonance phenomenon induced by fluid–structure interaction.
       	The modal decomposition indicates that resonance arises from the coupling between the first and second structural modes, where the excitation of the second structural mode plays a critical role.
		Due to the combined effects of added mass and periodic stiffness variations, the two modes become synchronized, oscillating at the same frequency while maintaining a fixed phase difference of $\pi/2$. 
		This further results in the resonant frequency being locked at half of the second natural frequency, which is approximately three times the first natural frequency. 
		A reduction in plate length and an increase in mass ratio are both associated with a narrower resonant locking range, while a higher mass ratio also shifts this range toward lower frequencies.	
		A symmetry-breaking bifurcation is observed for cases with $L/D\leq3.5$, whereas for $L/D=4.0$, the flow remains in a steady state with a stationary splitter plate prior to the onset of resonance. 
		For cases with a short flexible plate and a high mass ratio, the shortened resonance interval causes the plate to return to the symmetry-breaking stage after resonance, gradually approaching an equilibrium position determined by the flow field characteristics at high flexibility coefficients.

		\end{abstract}
		

{\bf MSC Codes }  {\it(Optional)} Please enter your MSC Codes here

\section{\label{sec1}Introduction}

Fluid–structure interaction (FSI) problems involving significant structural deformation are prevalent in aerospace, civil, and mechanical engineering applications \citep{chyu1996heat,defraeye2010cfd,kim2015proximity,cao2019investigation,CHEN2020104312}. 
In such flows, flow separation and alternating vortex shedding in the wake could lead to significant increases in mean drag and lift fluctuations, often resulting in flow-induced vibrations (FIV). These vibrations are undesirable in many engineering applications, since it can cause fatigue damage and catastrophic consequences \citep{nozawa2002large,zhao2024free,cao2022topological,song2022direct}. 
Additionally, flexible plates attached to bluff bodies have garnered significant attention in biomimetic studies, such as those modeling fish swimming and bird wing flight \citep{lim2007bluff}. In this study, we focus on a canonical FIV configuration: a flexible plate attached to a circular cylinder. We systematically investigate the vibration modes of the plate and the transitions between these modes.

Almost seven decades ago, rigid splitter plates attached to the rear of a circular cylinder were shown to significantly modify wake characteristics \citep{roshko1954drag, apelt1975effects, kwon1996control}. Subsequent experimental and numerical studies on cylinder-rigid splitter systems have extensively investigated the influence of various guide vane configurations on wake dynamics and vibration suppression. \citet{apelt1973effects} and \citet{apelt1975effects} pioneered research on splitter plate effects, demonstrating that vortex shedding is completely suppressed for plate lengths $L$ exceeding five times the cylinder diameter $D$ at Reynolds numbers $Re = 10^{4} \sim 5 \times 10^{4}$. The experimental results revealed a gradual decrease in vortex shedding frequency, accompanied by drag reduction as the plate length extends from $2D$ to $5D$.
\citet{kwon1996control} conducted a numerical investigation of splitter plates ($L/D=0\sim2$) attached to the base of a cylinder in laminar flow. 
They found that the critical length at which vortex shedding behind the cylinder completely ceased was proportional to the Reynolds number.
\citet{unal1988vortex} experimentally investigated the effect of the gap ratio between the cylinder and the splitter plate on the wake dynamics, categorizing the wake region into two distinct regimes: the pre-vortex formation region and the post-vortex formation region. An abrupt pressure variation on the splitter plate was observed when the plate was moved from the pre-vortex formation regime to the post-vortex formation regime. 

Recently, flexibility has been introduced into cylinder-splitter systems, greatly enriching their fluid--structure interaction dynamics. \citet{shukla2009flow} investigated the flow over a cylinder with a rigid splitter plate, which was allowed to rotate about a hinge point at the base of the cylinder. Symmetric periodic oscillations of the plate were observed, induced by the interaction between the cylinder wake and the plate. Additionally, a distinctive asymmetric motion with a lateral equilibrium shift was identified when the damping of the hinge fell within a critical range.
This deflection of the equilibrium position can be attributed to a symmetry-breaking phenomenon \citep{Crawfor1991}. 
Symmetry-breaking in fluid dynamics refers to the phenomenon where the flow loses its initial symmetry, leading to the emergence of asymmetric patterns or structures due to certain instabilities.\citep{cummins2018separated,alben2005coherent,vandenberghe2004symmetry,gadelha2010nonlinear}. 
This transition to different asymmetric bifurcations is usually affected by small perturbations in the flow and nonlinear interactions. 
\citet{assi2009low} also reported asymmetric displacement in a spring-supported cylinder attached with a rigid splitter plate.
They used different number of splitter plates with or without a gap to suppress the vortex-induced vibrations of circular cylinders. The results indicated that the attached single splitter plate developed a mean transverse force, which can be eliminated by using a dual splitter plate arrangement. Moreover, by varying the length of the splitter plate, \citet{shukla2009flow} classified the response of hinged-rigid splitter plates into two distinct regimes. For $L/D < 3.5$, the plates exhibit strong periodic oscillations, whereas for $L/D > 3.5$, the oscillations are weaker and characterized by a broad-band peak in their spectra, similar to the velocity spectra observed for long fixed–rigid splitter plate.%

Continuously deformable flexible plates attached to cylinders have garnered significant attention for their potential to enhance the propulsion efficiency of the cylinder, akin to the role of flexible fins or tail structures in fish and flexible wings in aircraft. The local pressure differences arising from shear layers on both sides of the plate can induce deformations, leading to more complex FIV behavior of the plate.
\citet{lee2013flexible} conducted a computational investigation into wake-induced vibrations of a flexible splitter plate attached to the base of a cylinder at a low Reynolds number. As the motion characteristics of a hinged plate are related to the damping of the hinge, the dynamic response of the flexible plate is determined by its stiffness. By varying the stiffness of the flexible plate, \citet{lee2013flexible} found that the Strouhal number of vortex shedding or deflection frequency of the plate cannot be easily predicted using the natural frequencies of the plate. 
The vibration characteristics of splitter plates exhibited significant mode dependence on aspect ratios. Plates with $L/D = 1$ demonstrated predominant first structural mode, whereas those with $L/D = 2$ primarily vibrated in second structural mode patterns. Notably, the longer splitter plate ($L/D = 3$) displayed hybrid modal behavior, with its deflection profile superimposing characteristics of both first two natural modes.
\citet{wu2014flow} presented a numerical analysis of flow over a fixed or elastically mounted circular cylinder with an attached flexible filament. The results suggested that shorter filaments attached to the fixed cylinder are prone to symmetry-breaking vibrations, while longer flexible filaments generally maintain a symmetric oscillation pattern unless their flexibility is sufficiently low to induce lateral deflections. For an elastically mounted cylinder, a long filament causes the oscillating cylinder to deviate from the wake centerline. 
\citet{sahu2019numerical} conducted a numerical study at $Re=150$ to investigate the
effects of splitter plate flexibility and length on vortex-induced vibration (VIV) and galloping instability. They identified lock-in with up to the second structural mode, with the vibration frequency of the plate tip in different lock-in regimes closely matching the natural frequency of the corresponding Euler–Bernoulli bending mode. 
By exploring a significantly broader range of splitter plate flexibility and length,  \citet{Furquan_Mittal_2021} observed lock-in up to the fourth structural mode and a variety of vortex-shedding modes. They also examined the response of the fluid–plate system at subcritical $Re$, where steady flow past a cylinder with a rigid plate remains stable. However, aeroelastic coupling destabilizes the system, leading to VIV and flutter even at subcritical $Re$.
\citet{pfister2020fluid} conducted a linear stability analysis of the coupled fluid–structure system, investigating the role of eigenmodes in the self-excited vibration of the plate.
A symmetry-breaking unstable mode revealing the underlying mechanism driving nonlinear oscillation bifurcations was found in small stiffness.
\citet{sahu2023symmetry} used steady-state computations to determine the equilibrium position of a flexible splitter plate of $3.5D$ attached to a fixed circular cylinder in a steady flow field, providing insight into the symmetry-breaking phenomenon.
However, the aforementioned studies on bifurcation dynamics used limited plate lengths, which restricts the ability to draw broad conclusions, as splitter length plays a significant role in the stability of the flow field.

\begin{figure}
	
	\centering
	\includegraphics[height=5.0cm]{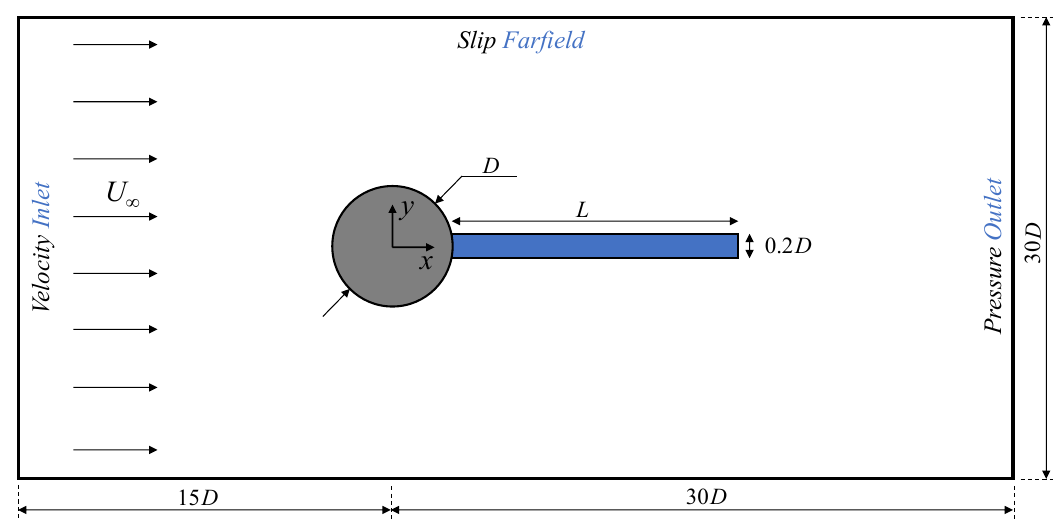}
	
	\caption{\label{fig1} Schematic view of the configuration: the flexible plate attached to circular cylinder immersed in the laminar flow. }
	
\end{figure}

It is well known that in a single-degree-of-freedom cylinder model supported by springs, when the vortex shedding frequency approaches the natural frequency of the structure, large-amplitude resonance occurs. However, in fluid–structure interaction problems involving flexible plates, predicting different vibration modes based on their higher-order natural frequencies is quite challenging. Despite numerous studies on flexible splitter plates attached to cylinders, the resonance mechanism responsible for the large-amplitude vibrations has yet to be fully elucidated. To address these gaps, the effects of the plate length $L/D$ and flexibility coefficient $S^{*}$ and mass ratio $m^{*}$ on the wake dynamics, plate vibrations, and flow stability are systematically explored. The unsteady response of the flexible plate and its interaction with the wake flow are expected to be clarified. Specifically, the key questions that we seek answers to are:
($a$) What is the underlying triggering mechanism of the large-amplitude resonance phenomenon, and how is it influenced by plate length and mass ratio?
($b$) How does vortex formation and evolution interact with a vibrating flexible plate, and what is its feedback on the vibration amplitude and vibration modes of the plate?
($c$) How do variations in plate length and symmetry-breaking phenomena affect the stability of the flow field, and what is its relationship with the resonance phenomenon?


\section{\label{sec2}Numerical Methodology and Simulation Details}

\subsection{Problem description and mathematical formulation}

As depicted in figure \ref{fig1}, the flow over a fixed circular cylinder attached with a flexible splitter plate of plate length $L$ ranging from $2.0D$ to $4.0D$, is investigated by numerical simulation. Here $D$ is the diameter of the leading cylinder. 
We work with a Cartesian coordinate with $x$ and $u$ in the horizontal (or streamwise) direction and $y$ and $v$ in the vertical (or wall-normal) direction.  
Based on the cylinder diameter $D$, a low Reynolds number of $Re = 150$ is considered in this work.  The non-dimensional parameter $Re$ is given by $Re = U_{\infty} D/\nu$, where $U_{\infty}$ is the inflow velocity. The quantity $\nu_{f}$ represents the kinematic viscosity of the fluid.
The time-varying response of the flexible plate is determined by the structural deformation equations.
The material properties of the plate are characterized by the Young modulus $E$, the Poisson ratio of $\nu_{s}=0.3$, and the mass ratio of $m^{*}=\rho_{s} / \rho_{f}$. Here, $\rho_{f}$ and $\rho_{s}$ are the density of fluid and solid, respectively. The flexibility coefficient $S^{*}$ proposed by \citet{sahu2023symmetry} is used to measure the flexibility properties of the material, which is defined as:
\begin{equation}
S^{*}=\frac{2 \pi}{1.875^{2}} \sqrt{\frac{12}{D^{2}}   \frac{\rho_{f} U^{2}_{\infty} L^{4}}{E H^{2}}     } 
\end{equation}
where,  $H$ represents the thickness of the plate, fixed at $0.2D$. The flexibility coefficient $S^{*}$ adopted in the present work ranges from 4 to 24, achieved by varying the normalized Young modulus $E/(\rho_{f}U_{\infty}^{2})$ from the maximum value 15331.9 at $L/D=4$ to the minimum value $26.6$ for the case with plate lengths $L/D=2$. For cases with a flexible plate length of $3D$, the influence of mass ratio on fluid-structure interaction is investigated for values of $m^{*}=1$, $10$, $20$, and $100$. For other cases of flexible plate length, the mass ratio is fixed to 10.

For this problem, the governing equations used to describe the fluid dynamics are the incompressible Navier-Stokes equations, which are written in non-dimensional form as follows:

\begin{equation} \label{eq1}
	\dfrac{\partial \boldsymbol{u}}{\partial t} +  (\boldsymbol{u} \bcdot \bnabla) \boldsymbol{u} = - \dfrac{1}{\rho_{f}} \bnabla p + \nu_{f}  \nabla^{2} \boldsymbol{u},
\end{equation}
\begin{equation} \label{eq2}
	\bnabla \bcdot \boldsymbol{u} = 0. 
\end{equation} 
where $\boldsymbol{u}=u_{i} (i=1,2)=(u, v)$ represent the velocity components with respect to the Cartesian coordinates $(x,y)$, $p$ is the pressure. 

Regarding the flexible splitter plate, the governing equation is the weak form of the balance of momentum, which is written in differential form as follows:

\begin{equation} \label{eq3}
	\rho_{s} \dfrac{\text{D} \boldsymbol{u}_{s}}{\text{D} t}  =  \bnabla \bcdot \boldsymbol{S},
\end{equation}

where $\boldsymbol{u}_{s}$ represents the velocity vector of the solid material point, $\boldsymbol{S}$ denotes the second Piola–Kirchhoff stress tensor. In present work, body force term is not included. The relationship between stress and strain is considered to be linear elastic. In order to have a set of closed-form equations, an equation that relates stress and strain is required,
\begin{equation} \label{eq4}
	\boldsymbol{S}  =  \text{C} : \text{E},  \text{E}=\dfrac{1}{2} (\text{F}^{\text{T}}  \text{F} - \delta).
\end{equation}
where $\text{C}$ is the elasticity tensor, $\text{E}$ represents the Green–Lagrange strain tensor, the symbol “:” represents the contraction of two tensors, $\text{F}$ denotes the deformation gradient, and $\delta$ represents the unit tensor.

The velocity and traction fields are continuous across the interface of the fluid and structure, $\Gamma_{fs}$:
\begin{equation} \label{eq5}
	\boldsymbol{u}_{f}  =  \boldsymbol{u}_{s} ,   ~ \sigma_{f} \cdot \vec{n} - \sigma_{s}\cdot \vec{n} = 0.
\end{equation}
Here, $\sigma_{f}$ and $\sigma_{s}$ denote the Cauchys stress tensor for the fluid and structure,
respectively, and $\vec{n}$ is a vector normal to the interface.

\subsection{\label{sec2.2}Computational simulation setup}
The secondary instability \citep{wang2019flow} in cylinder wake flow emerges at a Reynolds number of approximately $180$, with the flow field maintaining two-dimensionality below this critical threshold \citep{barkley1996three,williamson1996three}. Recent findings indicate that three-dimensional transition can be delayed both through cylinder transverse vibrations \citep{prasanth2008vortex,leontini2007three} and via rear splitter plate configurations.
Since the present study is restricted to $Re = 150$, a two-dimensional numerical simulation is conducted. 
The computational domain sizes in the streamwise direction and in the cross-flow direction are 45$D$, 30$D$, respectively. The fixed cylinder is placed $15D$ downstream from the inlet. The boundary conditions for the physical model are set as follows: For the circular cylinder surface, a no-slip velocity of $(u = v = 0)$ was specified, while the boundary condition for the flexible plate surface follows Eq.(\ref{eq5}). The far-field in the lateral flow direction employs slip boundary conditions. Besides, a Neumann boundary condition $(\partial u/\partial x = \partial v/\partial x = 0)$ is applied at the outlet. For pressure, a high-order Neumann type condition is implemented at the inlet and the wall, and the outlet uses a Dirichlet boundary condition ($p=0$).

The two way fluid-structure interaction problem is solved utilizing a block-iterative partitioned approach. 
The governing equations for the fluid dynamics (Eq.(\ref{eq1}),(\ref{eq2})) and structure deformation (Eq.(\ref{eq3})) are solved using open source framework \href{www.openfoam.org}{OpenFOAM} and \href{www.calculix.de}{CalculiX}, respectively. The fluid solver provides the loads that act on the structure at its interface with the fluid. The response of the structure to these loads, computed from the structure solver, alters the velocity and the shape of the interface, to be conformed by the fluid. The fluid flow and structural analyses are carried out alternately until convergence is achieved at each time step. Each of the fluid solver and structure solver is based on an implicit method.

For the fluid solver, direct numerical simulations with the moving-mesh technique are utilized to solve the flow field variables. Regarding the spatial discretization, the finite volume method (FVM) is adopted. Both the convection and diffusion terms are discretized at second-order accuracy with centred schemes. For the time advancement, we utilize the pressure implicit with splitting of operators (PISO) algorithm.  The time integration of Eq.(\ref{eq1}) is conducted using the second-order backward scheme. The geometric-algebraic multi-grid (GAMG) preconditioner \citep{behrens2009openfoam} is used for solving linear systems for the pressure, which are considered to be converged when the residuals are less than $1\times 10^{-6}$. The symmetric Gauss-Seidel method is used for solving linear systems for velocities with a local accuracy of $1\times 10^{-7}$. The finite element method was adopting to conduct computational structure dynamics problem. It utilizes the Galerkin method to solve Eq.(\ref{eq3}) and accounts for geometric nonlinearity. The semi-discrete equations are integrated in time by the modified NewMark scheme.

Two solvers are linked through the open-source coupling library preCICE \citep{preCICEv2}. The coupling procedure is configured with a parallel-implicit coupling scheme \citep{mehl2016parallel} using the IQN-ILS (interface quasi-newton with inverse Jacobian from least-square model) acceleration algorithm because of its robustness and good convergence. An implicit scheme is employed in the current FSI simulations, while the non-dimensional time step is set as the same value of 0.002 for the fluid and solid solvers. The radial-basis function (RBF) \citep{karayiannis2003construction} is adopted to interpolate and map the date of fluid–solid interface due to the difference in the surface meshes.

\subsection{Linear stability analysis}
A global direct stability analysis \citep{wang2019flow} is performed to assess the stability of the flow around a cylinder with an attached splitter plate.
The incompressible Navier–Stokes equations are linearized by decomposing the total flow states ($\boldsymbol{u},p$)
 into steady base states ($\overline{\boldsymbol{U}},\overline{P}$) and infinitesimal perturbations ($\tilde{\boldsymbol{u}},\tilde{p}$).
The base flow ($\overline{\boldsymbol{U}},\overline{P}$) represents the equilibrium configuration about which stability is examined.
While the perturbation terms are assumed to be in the form of normal modes $(\tilde{\boldsymbol{u}}(x,y,t),\tilde{p}(x,y,t))^{T} =  (\hat{\boldsymbol{u}}(x,y), \hat{p}(x,y))^{T}e^{\sigma t}$.
Substituting perturbation decomposition into Eq. (\ref{eq1}) and (\ref{eq2}), and neglecting the products of the perturbation fields, we obtain the linearized Navier–Stokes equations around the base state ($\overline{\boldsymbol{U}},\overline{P}$): 
\begin{equation} \label{eq6}
	\sigma \hat{\boldsymbol{u}} + (\hat{\boldsymbol{u} } \bcdot \bnabla)  \overline{\boldsymbol{U}} + (\overline{\boldsymbol{U}} \bcdot \bnabla  )\hat{\boldsymbol{u}} = - \dfrac{1}{\rho_{f}}\bnabla\hat{p} +  \nu_{f} \nabla^{2} \hat{\boldsymbol{u} }, 
\end{equation}
\begin{equation} \label{eq8}
	\bnabla \bcdot \hat{\boldsymbol{u} } = 0.
\end{equation}
The above linear perturbation equations govern the growth of perturbations to the leading order.
For the base flow, when the Reynolds number exceeds the critical $Re_{c}$, the cylinder wake undergoes a Hopf bifurcation \citep{duvsek1994numerical,agnaou2016steady}, transitions from a stationary, symmetric regime to one characterized by periodic oscillations. Consequently, the steady-state solution cannot be obtained directly. To address this, we employ the selective frequency damping (SFD) method developed by \citet{aakervik2006steady}, which suppresses unsteady temporal oscillations through the application of a low-pass filter. The boundary conditions for the base flow are set to be the same as those used in the unsteady simulations detailed in Section \ref{sec2.2}.
In accordance with the base flow boundary conditions, the perturbation velocity $\hat{\boldsymbol{u} }$
is set to zero at all boundaries, except at the outflow boundary, where a Neumann-type condition is imposed.
Let $\hat{q}$ to represent $(\hat{\boldsymbol{u}}(x,y),\hat{p}(x,y))^{T}$, the linear global modes of the equations in (\ref{eq6}) and (\ref{eq8}) can be obtained by solving an eigenvalue problem:
\begin{equation}
	\setlength{\arraycolsep}{0pt}
	\renewcommand{\arraystretch}{1.3}
	\sigma \hat{q} = \boldsymbol{A} \hat{q}, ~\text{with} ~
	 \boldsymbol{A} = \left[
	\begin{array}{cc}
		  - (\bnabla  \overline{\boldsymbol{U}}) - (\overline{\boldsymbol{U}} \bcdot \bnabla) + \nu_{f}\nabla^{2}   ~~~~   &     - \dfrac{1}{\rho_{f}}\bnabla    \\
		\displaystyle
		\bnabla \bcdot &  0   \\
	\end{array} \right].
	\label{defQc}
\end{equation}
As one can see, the Jacobian matrix $\boldsymbol{A}$ depends on the base state ($\overline{\textbf{U}},\overline{P}$).  The above equation can be solved by using a Krylov subspace method together with the Arnoldi iteration algorithm \citep{mamun1995asymmetry,barkley1996three,carmo2008wake}. By generating an orthogonal basis for Krylov subspace $K_{m}$, the large-scale eigenvalue problem for the matrix $\boldsymbol{A}$ is projected onto a smaller one of Hessenberg form that can be efficiently solved \citep{arnoldi1951principle,saad1980variations}. 
The stability of the system, characterized by the growth or decay of perturbations, is determined by the signs of the real parts of the eigenvalues, $Re(\sigma_{i})$, which are referred to as the growth rates of the unstable modes.
While the imaginary part $Im(\sigma_{i})$ represents the frequency.
The eigenvectors (or eigenmodes) corresponding to large or positive $Re (\sigma_{i})$ offer insight into the spatial structure of the most unstable disturbances.

\subsection{Computational mesh details and validation}

\begin{table}
	
	\def~{\hphantom{0}}
	\centering
	\begin{tabular}{cccccccc}
		Case & Number of grid elements & $h^{+}$ & $St$ & $A_{tip}$ & $Y_{tip}(S^{*}=8.05)$ \\ 
		\hline
		Coarse & $8.2 \times 10^{4}$ & $ 0.01D$  &$0.132$  &$0.831D$  &$0.181D$      \\ 	
		
		Medium & $12.6 \times 10^{4}$ & $ 0.005D$  &$0.132$  &$0.971D$  &$0.161D$    \\ 
		Fine   & $15.3 \times 10^{4}$ & $ 0.005D$  &$0.133$  &$0.977D$  &$0.158D$    \\
	\end{tabular}
	\caption{\label{tab1} Summary of the grid dependence test.  }
\end{table}
\begin{figure}
	\centerline{\includegraphics[height=4.5cm]{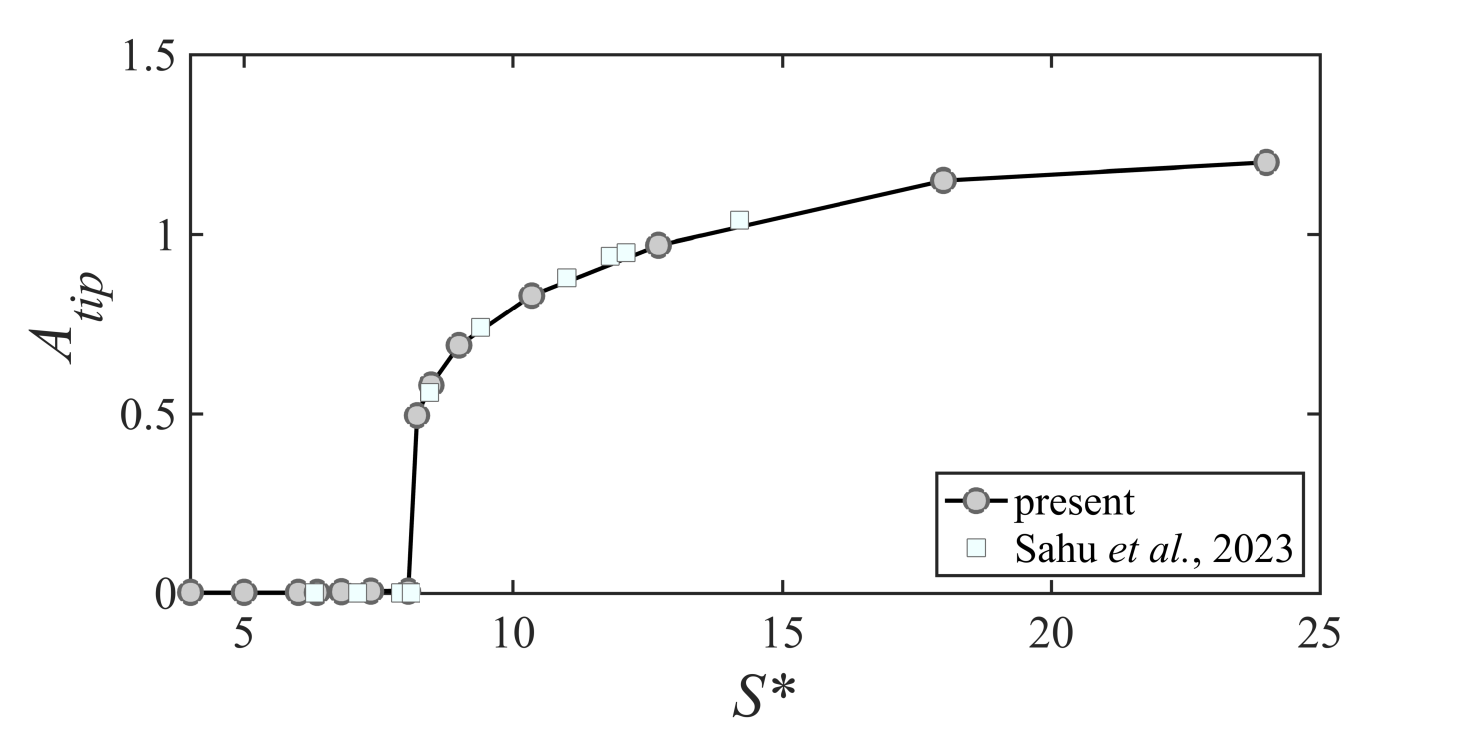 }}
	\caption{Variations of the amplitude of vertical displacement at the plate-tip against the normalized bending stiffness $S^{*}$ for the case with $L/D=3.5$.}
	\label{fig2}
\end{figure}

Both the flow field and the solid field are meshed using structured grids. The thickness of the first layer mesh near the cylinder surface $h^{+}$ can be given empirically as $ 0.1 / \sqrt{Re}$ \citep{cao2014velocity}, which is $0.001D$, and it is guaranteed that the maximum $Y^{+}$ is less than 1, while coarser mesh layers are applied progressively away from the solid surface. 
A grid sensitivity analysis is performed using the hexahedral mesh by varying the grid resolution for the case of a flexible splitter plate with $L/D = 3.5$ and $S^{*} = 18$. The flexible splitter plate is discretized using about $1500$ eight-noded elements with maximum size no more then $0.02D$ for the structural model. Results are examined for Strouhal number $St$,  amplitude of the vertical displacement $A_{tip}$ and mean displacement $Y_{tip}$ at the tip of the splitter plate, as shown in Table \ref{tab1}. 
All these global quantities have shown that medium mesh is fine enough for the convergence of the solution at $Re = 150$. By applying the same grid size control parameters to all kinds of $L/D$ cases, the total number of grids amounts to approximately $1.1 \times 10^{5}$ and $1.4 \times 10^{5}$ for the case with $L/D=2$ and $L/D=4$, respectively.

We further demonstrate the accuracy of our numerical method by comparing the results for the variation in amplitude of the vertical displacement $A_{tip}$, with those from numerical studies by \citet{sahu2023symmetry} under same plate length. As figure \ref{fig2} illustrates, the $A_{tip}$ distributions with changes in characteristic flexibility $S^{*}$ agree very well with each other. This result suggests that the present numerical method provides reliable results for the flow structure interaction dynamics under investigation.

\section{\label{sec3}Dynamic response of the cylinder with an attached flexible plate}
\subsection{\label{sec3.1}Vibration modes of the attached flexible plate with different plate length}

\begin{figure}
	
	\centerline{
		\includegraphics[height=4.5cm]{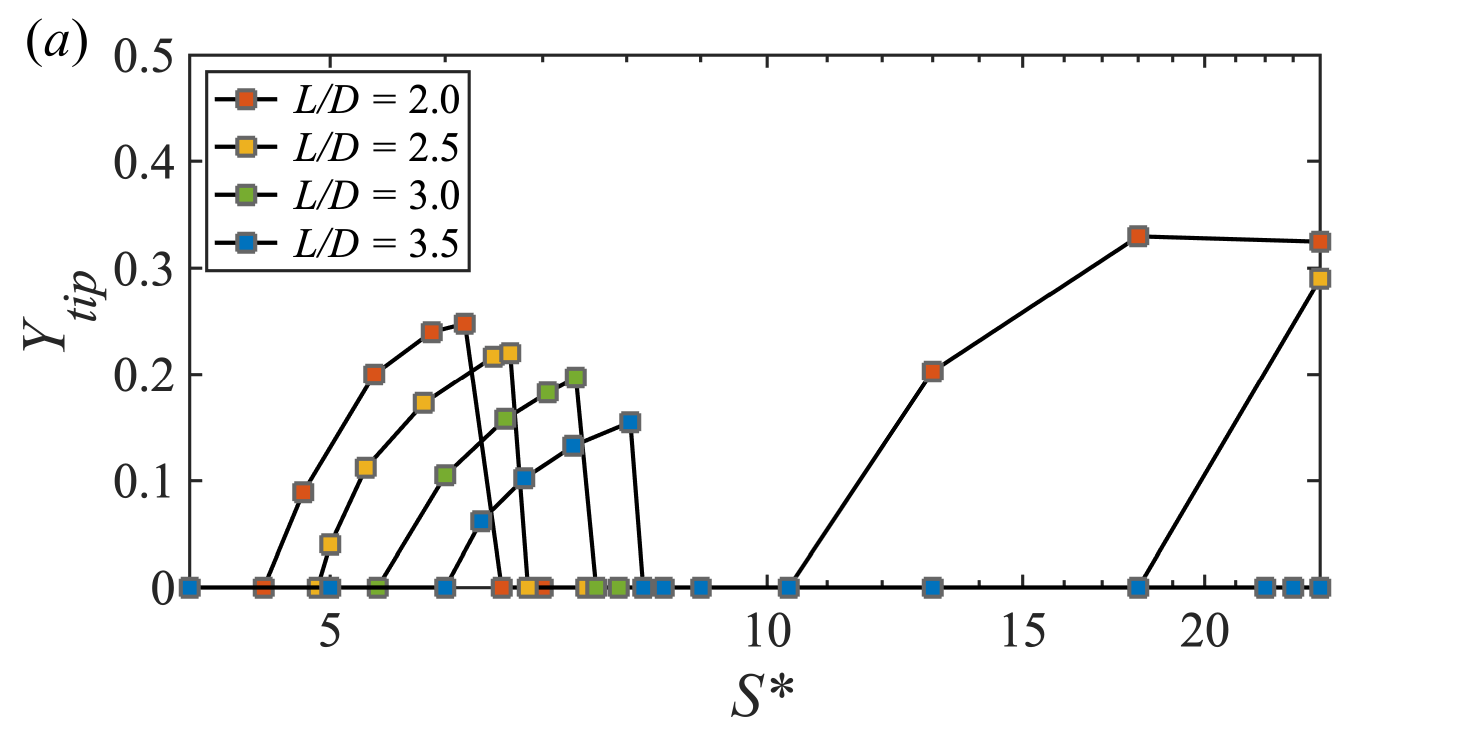}
		\label{fig3a}
	}
	\centerline{
		\includegraphics[height=4.5cm]{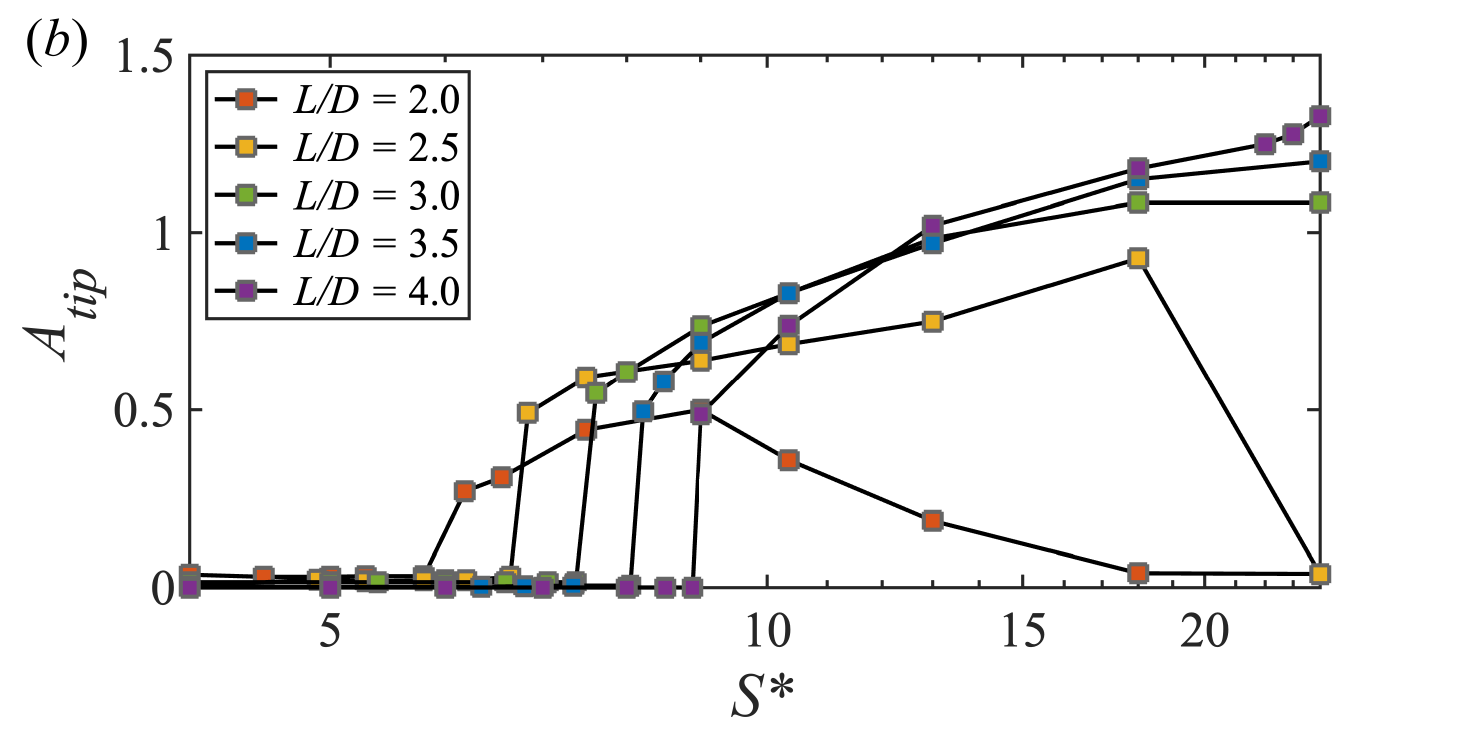}
	}
	
	\caption{ Oscillation characteristics of the flexible plate: variation of non-dimensional ($a$) the time-averaged position $Y_{tip}$ and ($b$) amplitude $A_{tip}$ of vertical displacement at the plate tip, with increasing flexibility coefficient $S^{*}$ for the cases with different plate length at $m^{*}=10$.}
	\label{fig3}
\end{figure}

The interaction between the shear layers separating from the bluff body and the flexible splitter plate exhibits highly nonlinear behavior. For the flexible splitter plate, the vibration modes vary depending on its material stiffness. By setting monitoring points at the free end of the plate and analyzing the displacement time histories, we can categorize its dynamic response into different types.
Figure \ref{fig3} shows the statistical parameters of the endpoint motion of the flexible plate under different length $L$ and flexibility coefficient $S^{*}$ at $m^{*}=10$.
The vertical mean displacement $Y_{tip}$ is related to the symmetry-breaking phenomenon. When the vertical mean displacement $Y_{tip}$ equals zero, it indicates that the splitter plate either remains stationary or undergoes symmetric vertical oscillations. When the mean displacement is non-zero, it means that the symmetry-breaking phenomenon has occurred. Figure \ref{fig3}($a$) shows that, except for the cases with an $L/D$ of 4, the symmetry-breaking phenomenon occurs in all other cases. In the symmetry-breaking bifurcation stage, as flexibility coefficient increases, the time-averaged position progressively deviates from the wake centerline. However, as the flexible plate becomes increasingly flexible, a transition occurs: the oscillation amplitude of the solid structure rapidly grows, while its mean displacement returns to zero. This transition signifies the end of the symmetry-breaking stage, with the solid structure shifting into a symmetric flow-induced vibration state. For cases with a splitter plate length of $4D$, the mean displacement remains zero across all $S^{*}$ values. However, the maximum amplitude variation curves in figure \ref{fig3}($b$) show that, prior to the sudden amplitude increase, the response amplitude remains at zero. This behavior suggests the occurrence of a similar transition phenomenon, consistent with that observed for other plate lengths.
In the $L=4D$ cases, the flow field remains steady at lower values of the flexibility coefficient, and no vibrations are observed in the flexible plate. As the flexibility coefficient increases, this steady state is disrupted once $S^{*}$ reaches 9. For symmetric vibrations at higher flexibility, the vibration amplitude increases with further growth in $S^{*}$.
In contrast, for $L=2D$, when $S^{*}$ exceeds 9, the trend reverses: the tip amplitude decreases as flexibility increases, accompanied by the reappearance of the symmetry-breaking phenomenon. A similar return to the symmetry-breaking state is also observed for $L=2.5D$ at $S^{*}=24$.

\begin{figure}
	\centerline{\includegraphics[height=4.5cm]{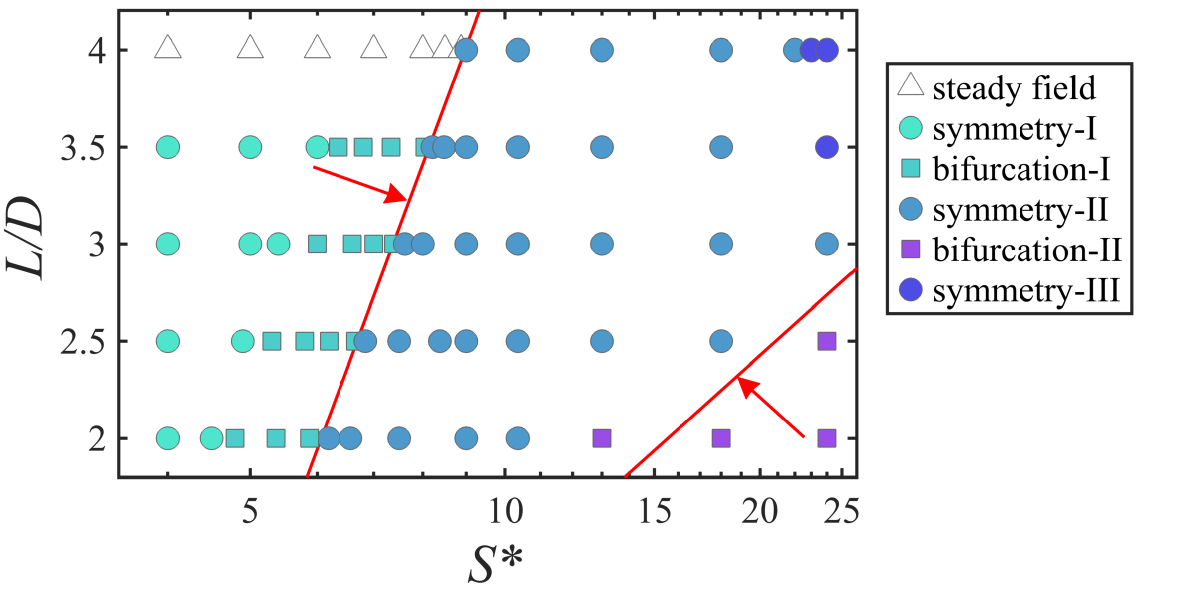}}
	\caption{Partition of the different vibration modes in the $L/D-S^{*}$ diagram. The region enclosed between the two red lines represents the parameter range in which large-amplitude responses occur. The lower-right corner of the parameter map shows the reappearance of the symmetry-breaking phenomenon at high flexibility parameter values. }
	\label{fig4}
\end{figure}

\begin{figure}
	
	\centerline{
		\includegraphics[height=2.0625cm]{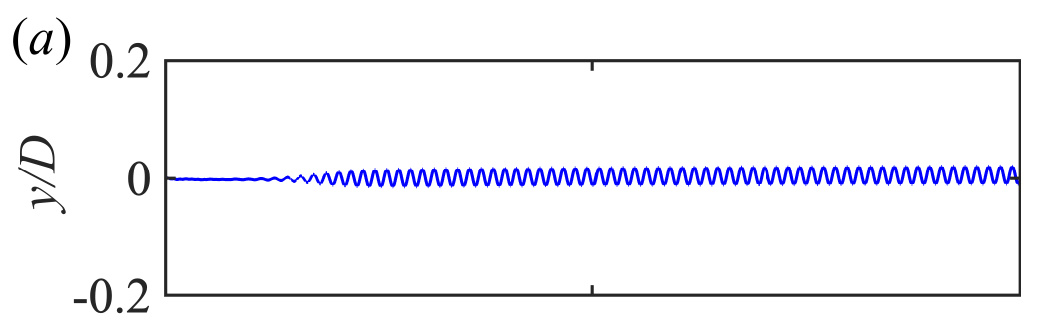}
		
		\includegraphics[height=2.0625cm]{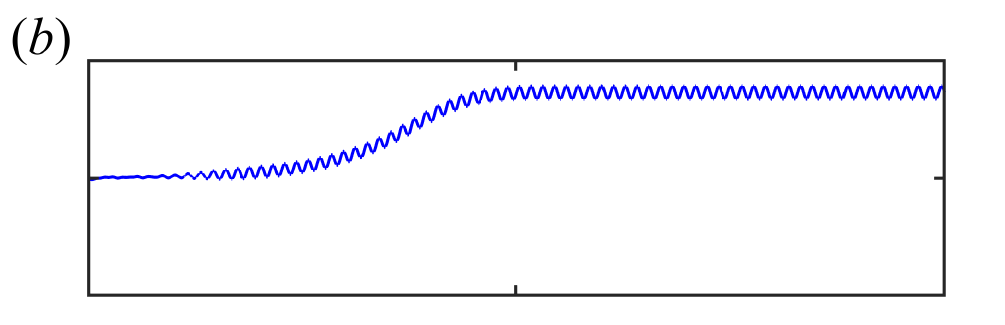}
	}
	\centerline{
		\includegraphics[height=2.625cm]{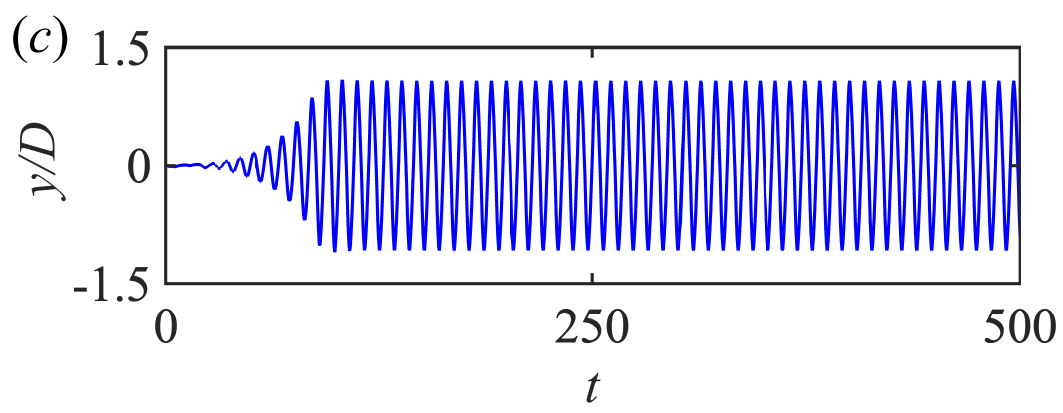}
		
		\includegraphics[height=2.625cm]{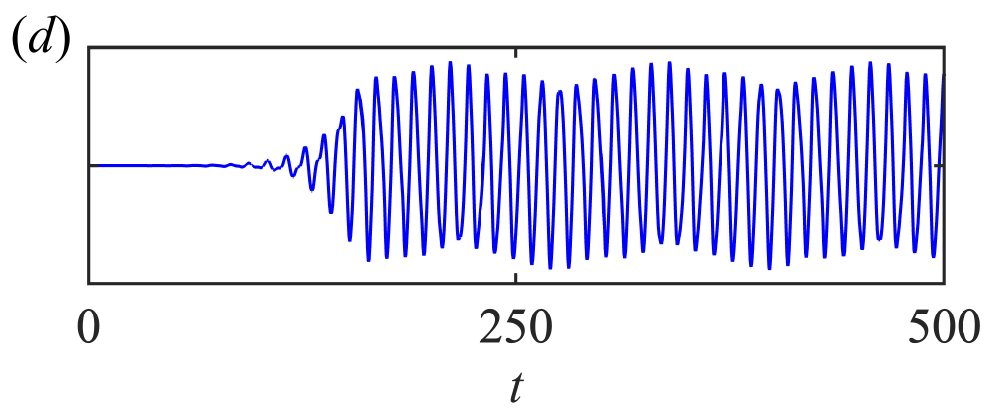}
    	}
	
	\caption{Time histories of the vertical displacement at the plate tip at different vibration modes: ($a$) symmetry-I stage for $S^{*}=5$ and $L/D=3$, ($b$) bifurcation-I stage for $S^{*}=7$ and $L/D=3$, ($c$) symmetry-II stage for $S^{*}=18$ and $L/D=3$, ($d$) symmetry-III stage for $S^{*}=24$ and $L/D=4$.}	
	\label{fig5}
\end{figure}

Figure \ref{fig4} summarizes the different vibration modes for cases with varying lengths $L$ and flexibility coefficient $S^{*}$ of splitter plates under fluid-structure interaction.
Cases in which the flexible plate undergoes symmetric vibration about the wake centerline are represented by circular markers, whereas those in the symmetry-breaking bifurcation stage are denoted by square markers. Different vibration types are further distinguished by color. For the stationary plate surrounded by a steady flow field at $L=4D$, white triangular markers are used. The symmetry-breaking phenomenon is categorized into two stages, bifurcation-I and bifurcation-II, based on the different flexibility ranges in which they occur.
For symmetric vibrations, the first stage corresponds to small-amplitude vibrations occurring before the onset of symmetry-breaking bifurcation, while the second stage corresponds to large-amplitude vibrations. 
The third stage of vibration differs significantly from the first two stages, as the vertical displacement time history of the plate tip exhibits quasi-periodic behavior. Figure \ref{fig5} illustrates the vertical displacement time histories of the plate tip in the different vibration modes. Figures \ref{fig5}$(a)$-$(c)$ show the time history curves of the transverse displacement at the plate tip for the cases with flexible plate $L/D=3$ during the symmetry-I, bifurcation-I, and symmetry-II stages, respectively.
Figure \ref{fig5}($d$) highlights the quasi-periodic characteristics observed during the third symmetric vibration stage. Such quasi-periodic characteristics were also observed in the work of \citet{pfister2020fluid} at a Reynolds number of 80. In the current numerical simulations, this phenomenon is only triggered in cases where the plate is relatively long and exhibits high flexibility. This quasi-periodicity is, in fact, associated with low-frequency wake oscillations, which will be discussed in detail in subsequent sections. Another interesting finding is that the reappearance of symmetry-breaking (bifurcation-II stage) as increasing flexibility coefficient exceeds a relatively high value after the symmetry-II stage. These cases are donated by purple square markers and are only found in the case with shorter plate length ($L/D=2, 2.5$).

\subsection {Flow pattern of the cylinder with an attached flexible plate}

\begin{figure}
	\centerline{\includegraphics[width=13.5cm]{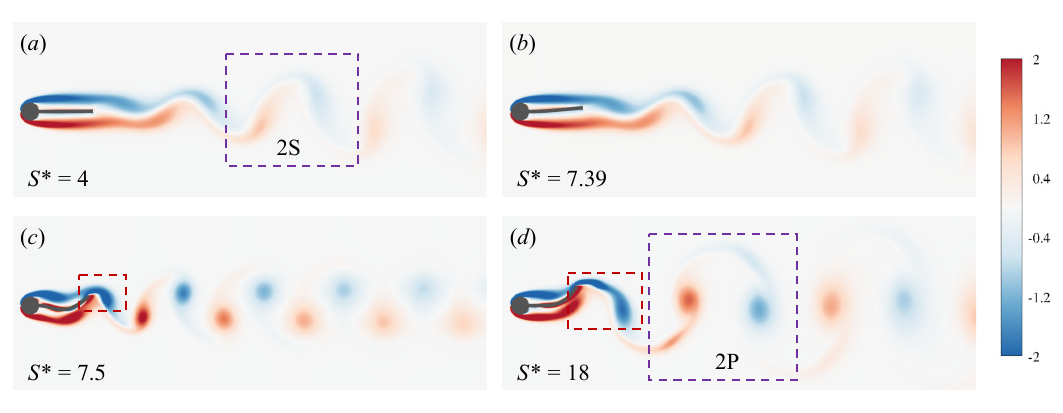}}
	\caption{Flow patterns visualized by spanwise vorticity $\omega_{z}$ for the case with the plate length of $ L = 3D (m^{*}=10)$ in different vibration stages: ($a$) symmetry-I stage, ($b$) bifurcation-I stage, ($c$) onset of the symmetry-II stage, ($d$) large flexibility coefficient case in the symmetry-II stage. 2S modes and '2P' mode are observed. The red dashed box highlights the interaction between the tail of the flexible plate and the cylinder wake vortex.	}
	\label{fig6}
\end{figure}

\begin{figure}
	\centerline{
		 \includegraphics[height=4.5cm]{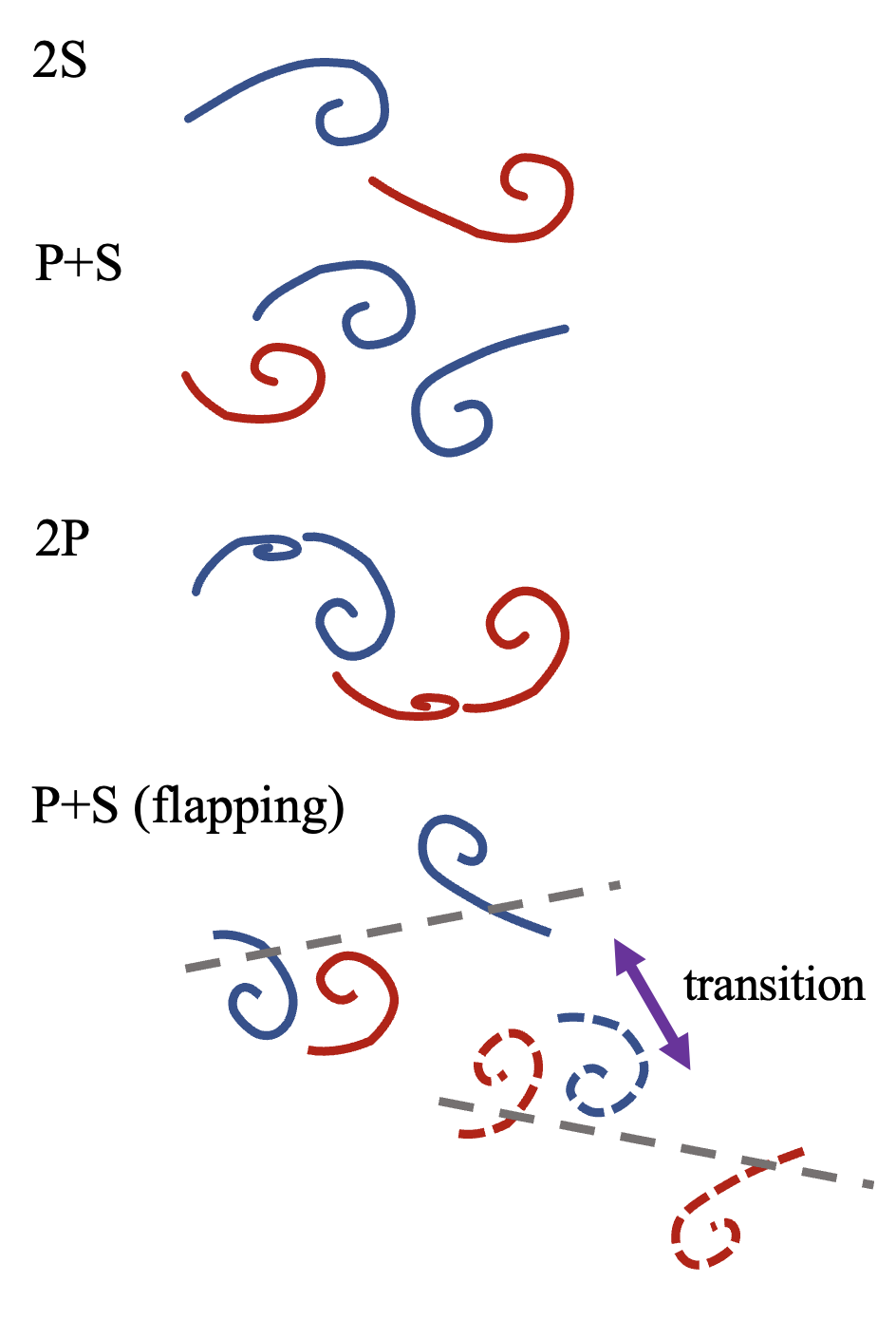}
		\includegraphics[height=4.5cm]{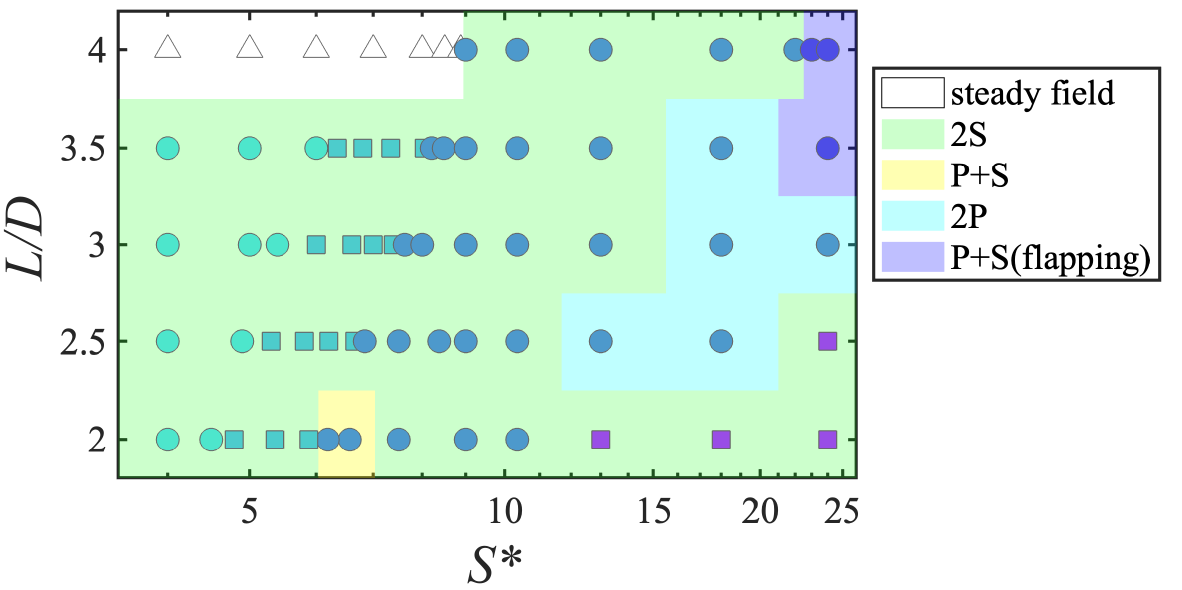}
	}
	\caption{Vortex-shedding pattern based regime map, for various plate length $L/D$ and normalized flexibility coefficient $S^{*}$, overlapped with different vibration modes. A distinct flapping P+S wake pattern has been identified, which is associated with quasi-periodic vibration.}
	\label{fig8}
\end{figure}

Oscillations of the flexible plate are accompanied by interactions with the cylinder wake, leading to changes in the wake dynamics. We further examined how the flow pattern varies with different plate lengths and changes in flexibility coefficient.
Figure \ref{fig6} presents contour plots of spanwise vorticity $\omega_{z}$ for the cases with the plate length of $ L = 3D $ at selected flexibility coefficient $S^{*}$. The flow field snapshots are taken at the moments when the vertical displacement of the flexible splitter plate tip reaches its maximum value. Through the spanwise vorticity contour plots, we can briefly summarize the fluid-structure interaction characteristics across the three stages. 
When the flexibility coefficient is small (figure \ref{fig6}($a$)), both the intensity of the vortex shedding in the wake and the frequency of vortex shedding are reduced with respect to the single cylinder case \citep{sahu2019numerical} due to the stabilizing effect of the splitter plate on the wake. And in this symmetry-I stage,  the wake exhibits a 2S mode (a pair of vortices is alternately shed per cycle) of flow pattern.
As the flexibility coefficient $S^{*}$ increases to $7.39$, marking the largest flexibility coefficient case in the bifurcation-I stage, the time-averaged vertical displacement of the plate tip nearly reaches its maximum value.
In the bifurcation-I stage, the vortex shedding pattern remains almost identical to that shown in figure \ref{fig6}($a$), except that the flexible plate tilts slightly to one side and undergoes minor oscillations synchronized with the vortex shedding. 
Figures \ref{fig6}($c$) and ($d$) show the flow fields around the cylinder with an attached flexible plate in the symmetry-II stage. The case shown in Figure \ref{fig6}($c$) represents the first instance after the transition from the symmetry-breaking state to the symmetry-II stage.
It can be observed that, although the flow pattern still shows a 2S mode, the interaction between the fluid and the solid enhances the strength of the wake vortices. 
As the flexibility increases further, two pairs of vortices shedding (2P) are presented in the wake, as shown in the figure \ref{fig6}($d$). However, the present two pairs of shed vortices in the 2P mode correspond to each pair shed by the cylinder and the plate as compared to the two pairs inverse vortex shed by the cylinder in the proposition by \citet{williamson1988vortex}.
As demonstrated in the red dashed boxes in figures \ref{fig6}($c$) and ($d$), in the case of larger flexibility coefficient in the symmetry-II stage, the vortices caused by the displacement of the splitter plate do not merge with the vortices shed from the main cylinder. Instead, this results in the formation of two large vortices each carrying two smaller vortices in the wake.
Moreover, the increase in the flexibility coefficient also causes the centers of the wake vortices to move closer to the wake centerline.

\begin{figure}
	\centerline{\includegraphics[width=13.5cm]{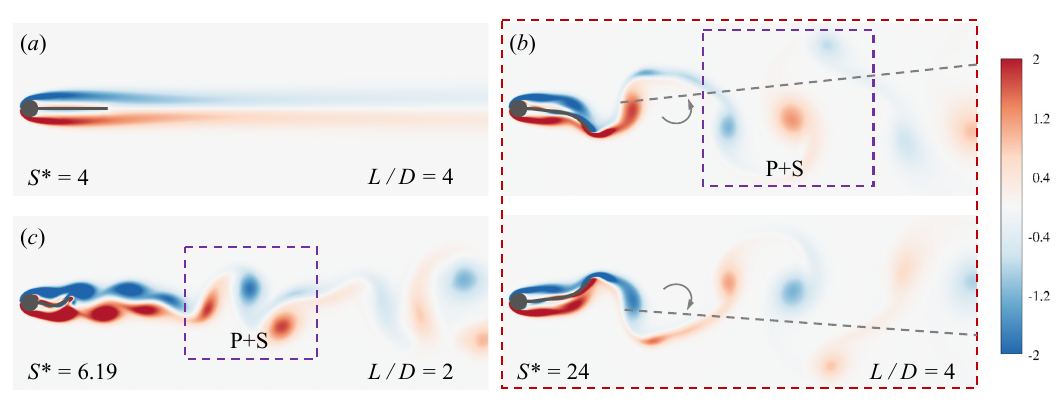}}
	\caption{Flow patterns visualized by spanwise vorticity $\omega_{z}$: ($a$) steady field, ($b$) flapping P+S mode, ($c$) stable P+S mode. The dashed gray line represents the centerline of the swinging wake.}
	\label{fig7}
\end{figure}
The distribution of wake modes under varying plate lengths and flexibility coefficients is presented in figure \ref{fig8}. 
In addition to the 2S mode and 2P mode of the wake shown in figure \ref{fig6}, figure \ref{fig7} illustrates three other flow pattern modes. When the flexible plate length is \( 4D \) and the flexibility coefficient is low, the stabilizing effect of the splitter plate results in a steady-state flow field that is symmetric along the wake centerline, as shown in figure \ref{fig7}($a$). Figures \ref{fig7}($b$) and ($c$) present two other P+S (a pair of vortices shedding from one side and a single vortex shedding from the other side) modes. The stable P+S mode typically occurs with shorter plate lengths and immediately after the symmetry-breaking stage, while the oscillatory P+S mode occurs with longer plate lengths under conditions of high flexibility coefficient $S^{*}$. 
As mentioned earlier, the quasi-periodic vibration of the flexible separation plate is associated with the vertical oscillation of wake vortex shedding.







\section {\label{sec5} Resonant response mechanisms in fluid–structure interaction}
\begin{figure}
	
	\centerline{
		\includegraphics[width=13.5cm]{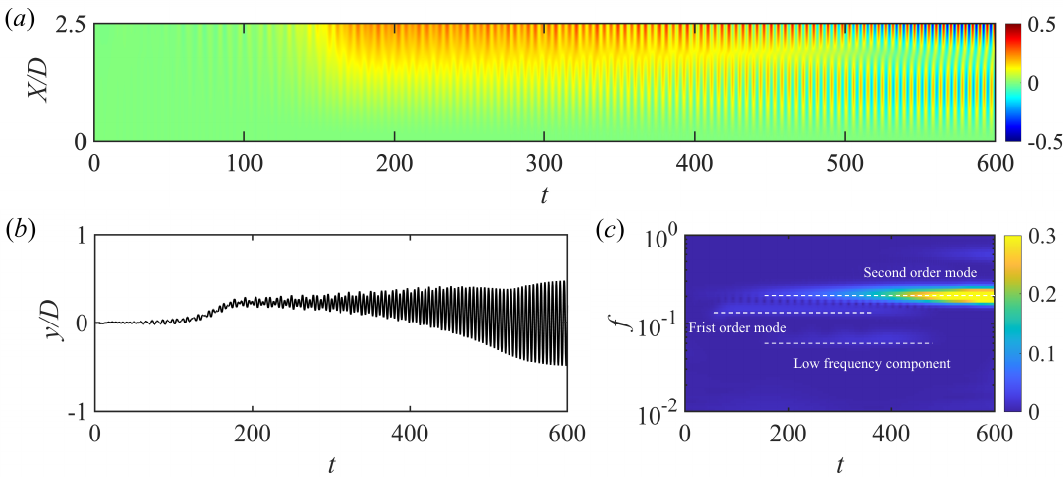}
	}
	
	\caption{Displacement variation of the flexible plate through the transition from bifurcation-I to symmetry-II stage: ($a$) Contour plot of transverse displacement $d_{y}(t,X)$ through non-dimensional time units $[0, 600]$ along plate length $[0, L]$ for the splitter plate with $L/D=2.5$ at $S^{*}=6.84$, along with the corresponding ($b$) time history of the vertical displacement at the plate tip and its ($c$) time varying frequency distribution.}
	\label{fig15}
\end{figure}
\subsection {Subharmonic lock-in at half the second natural frequency}

Immediately after transitioning into the symmetry-II stage, the system passes through a symmetry-breaking state before gradually destabilizing into a large-amplitude vibration regime.
Figure \ref{fig15} illustrates the transverse displacement variation over time unit $[0,600]$ for the case of plate length $L=2.5D$ at the lowest flexibility coefficient $S^{*}=6.84$ in the symmetry-II stage.
The contour plot of the time variation of the transverse displacement $d_{y}(t,X)$ along plate length $[0, L]$ (figure \ref{fig15}($a$)) shows that the flexible plate initially deflects to one side before gradually becoming unstable and transitioning to large-amplitude symmetric vibrations. 
This change can also be observed in the corresponding time history of the vertical displacement at the plate tip (figure \ref{fig15}($b$)).
When the flexible plate is still in the symmetry-breaking stage, the structural displacement is primarily concentrated at the trailing end, consistent with the shape of the first vibration mode.
As the amplitude at the tip of the flexible plate gradually increases, figure \ref{fig15}($a$) shows that vibrations with a phase difference relative to the tip vibrations begin to emerge in the middle of the plate.
This pattern with a phase difference of about 180 degrees confirms the appearance of the second structural mode.
Additionally, a region with smaller amplitudes exists between the vibrating areas at the tip and the middle of the plate, a vibration pattern that distinctly exhibits characteristics of the second vibration mode.
From figure \ref{fig15}($b$), it can be seen that, unlike regular periodic vibration curves, the vibration of the flexible plate contains distinct components of different frequencies, which gradually develop into large-amplitude symmetric vibrations.
Figure \ref{fig15}($c$) further presents a contour map of the time-frequency variation of the vertical displacement at the tip of the flexible plate, obtained through wavelet analysis. 
It can be observed that as the amplitude gradually increases, the frequency component representing the first structural mode diminishes, while a higher frequency of the second structural mode progressively develops. 
At the same time, during the conversion of the two vibration modes, a low-frequency signal appears.

\begin{figure}
	
	\centerline{
		\includegraphics[width=10cm]{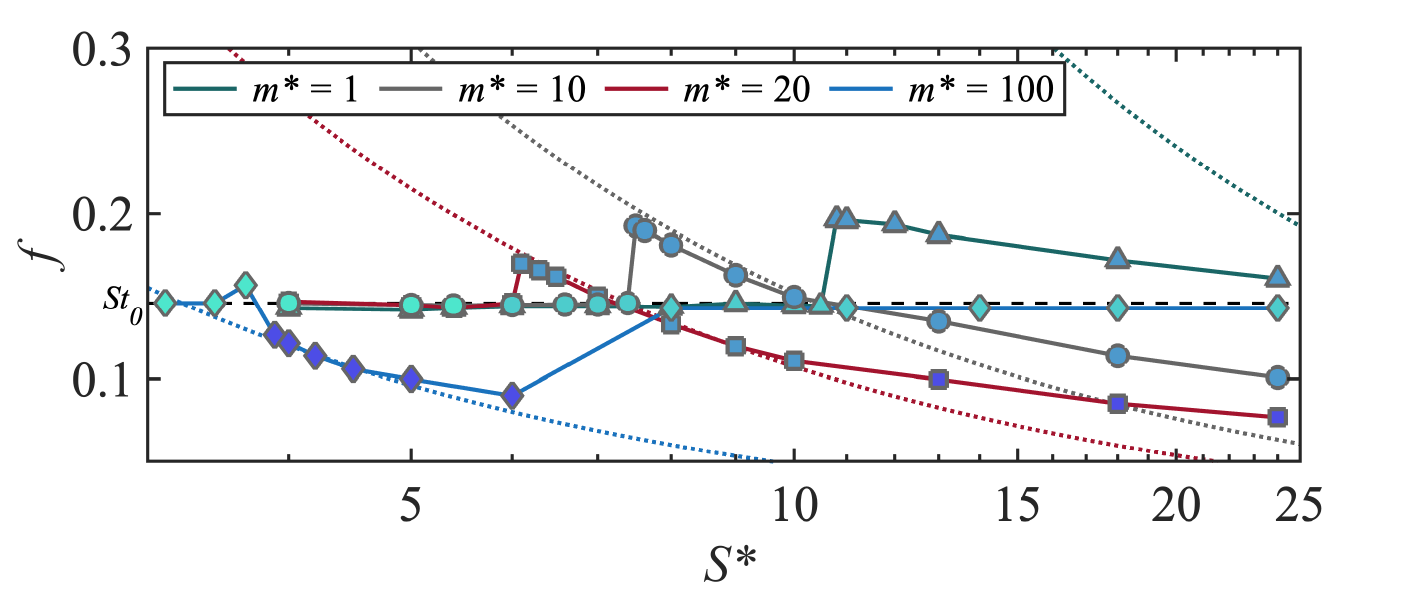}
	}
	
	\caption{Variation of the dominant vibration frequency and natural frequencies $f^{*}_{i}$ of the flexible splitter plate with $L/D = 3$ under different mass ratios $m^{*}$ as a function of the flexibility coefficient $S^{*}$. The dotted lines represent the subharmonic (half value) of the second natural frequency corresponding to each $m^{*}$ with same color.
The base vortex shedding frequency $St_{0}$, obtained from the rigid plate case, is indicated by the white dashed line.
Different vibration mode regimes are distinguished by marker colors.	
	}
	\label{fig16}
\end{figure}

When the flexible plate enters the symmetry-II stage, we can find in its displacement variation that the second structural mode component is gradually excited and transformed into a mode of large-amplitude motion accompanied by an increase in frequency. Such frequency variation is commonly attributed to resonance phenomena induced by fluid–structure interaction. In engineering applications, structural natural frequencies are typically designed to avoid coincidence with common excitation frequencies to ensure safety. In contrast, energy harvesting devices aim to exploit such resonance to enhance energy collection efficiency. To further investigate this resonance phenomenon, we analyze the frequency evolution of flexible plates with $L/D=3$ under different mass ratios $m^{*}$. Figure \ref{fig16} shows a point-line plot of the vertical displacement frequency at the splitter plate tip as a function of flexibility coefficient $S^{*}$. In order to obtain accurate natural frequency $f^{*}_{i}$, the eigenvalue problem obtained by the finite element method is adopted to solve the natural vibration mode, fully accounting for the two-dimensional thickness of the flexible plate. The same mesh and assumptions employed in the solid dynamics solver are used for these calculations. From the figure \ref{fig16}, it can be observed that when the flexibility coefficient of the splitter plate is low, the vibration frequency of the structure is close to the base vortex shedding frequency $St_{0}$ for the case with rigid splitter plate, indicating that the structure is primarily driven by the shedding of trailing vortices. As the flexibility coefficient increases, all cases with different mass ratios enter a large-amplitude vibration stage, accompanied by a noticeable change in vibration frequency. For flexible plates with a mass ratio $m^{*}\geq 10$, the observed frequency transition tends to lock onto half of the second natural frequency ($f^{*}_{2}/2$), with higher mass ratios resulting in frequencies more closely approaching this value, as denoted by the dotted line. For the flexible plate with a mass ratio of 1, the frequency deviation is relatively large, which may be attributed to the stronger influence of fluid-added mass in this case. With increasing mass ratio, another observed phenomenon is that the frequency-locking region associated with increased frequency gradually shifts toward the region of decreased frequency. When the dotted line decreases with increasing flexibility to a value smaller than the base vortex shedding frequency $St_{0}$, the vibration frequency of the structure becomes significantly higher than half of the second natural frequency, deviating toward the base vortex shedding frequency.

\subsection{Coupled resonance phenomenon of first and second structural modes}

\begin{figure}
	\centerline{\includegraphics[width=13.5cm]{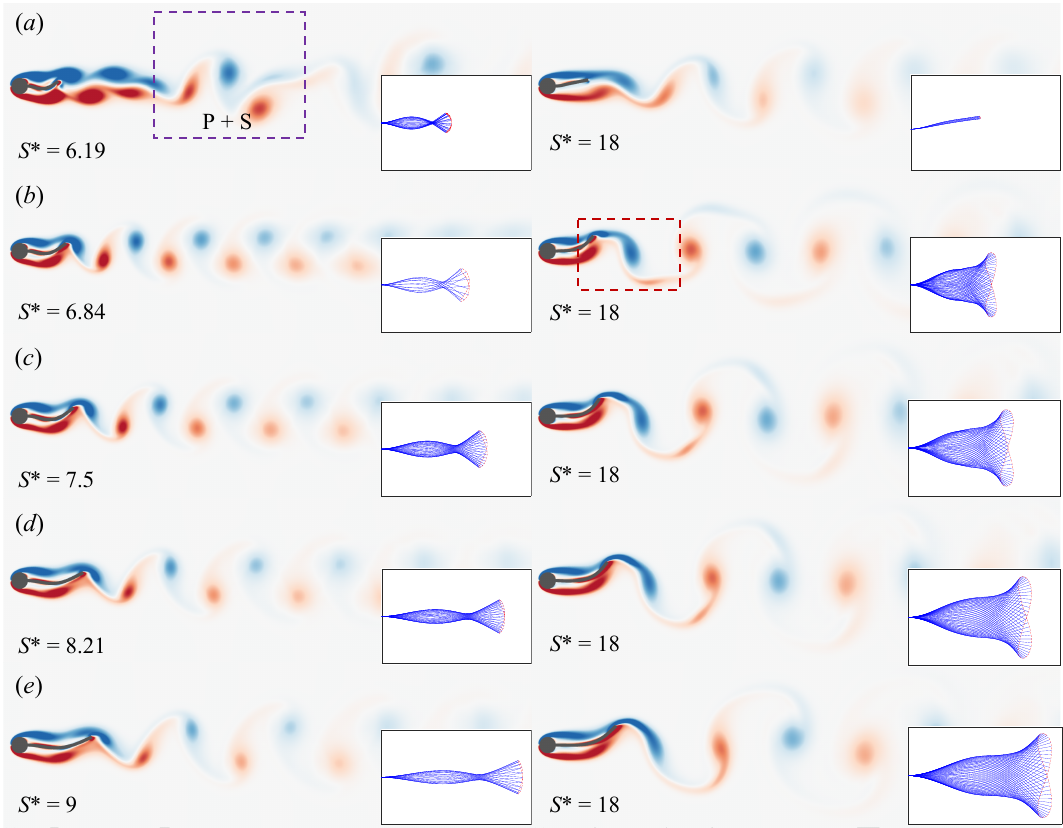}}
	\caption{Flow pattern visualization by spanwise vorticity $\omega_{z}$ contour plots and vibration
trajectory plots of the flexible plate for cases with $m^{*}=10$ different plate lengths: ($a$) $L/D=2$, ($b$) $L/D=2.5$, ($c$) $L/D=3$, ($d$) $L/D=3.5$, ($e$) $L/D=4$ The selected flexibility value
corresponding to the onset of transition in symmetry-II stage and a flexibility coefficient of $S^{*}=18$. The spanwise vorticity $\omega_{z}$ is limited in a range from $-2$ (blue) to $2$ (red).}
	\label{fig191}
\end{figure}

Figure \ref{fig191} presents spanwise vorticity $\omega_{z}$ contour plots and vibration trajectory plots of the flexible plate for cases with different plate lengths at $m^{*}=10$ and selected flexibility coefficient.
The vorticity plots are based on flow field snapshots taken at the moments when the plate tip displacement reaches its maximum value.
In the contour, the magnitude of the spanwise vorticity $\omega_{z}$ is limited in a range from $-2$ (blue) to $2$ (red).
The first column of contour plots in figure \ref{fig191} shows the cases at the onset of the symmetric-II stage. It can be observed that at the onset of the symmetric-II stage, the wake pattern for most cases is in the 2S mode, while for the case with the shorter plate length $L=2D$, the wake pattern is in the P+S mode.
For the vorticity contour plots in the first column, an obvious change is that as the plate length increases, the longitudinal distance between the wake vortices becomes longer.
The longitudinal distance reflects the frequency of vortex shedding, with longer longitudinal distances of vortices representing a lower frequency of vortex shedding.
To compare the effect of increased flexibility on the response dynamics, the larger flexibility coefficient $S^{*}=18$ cases are shown in the second column.
The lateral distance between vortices decreases for increased flexibility of flexible plates with plate lengths larger than $2D$.
This is reflected in the wake pattern by the alternating rows of vortices on the wake centerline.
Another noteworthy point is that for the case possessing less flexibility, the negative vortex generated by the upward swing of the flexible plate tip merges with the vortex separated from the cylinder wall. For the case of $S^{*}=18$, on the other hand, there appears to be a phase difference between the vortex shedding and the movement of the plate tip, with the vortex generated by the movement of the plate tip following the main vortex.
When we turn to observe the displacement trajectory of the flexible plate under different parameters, it is not difficult to find that the vibration mode of the flexible plate with less flexibility in this stage is closer to the second structural mode.
As for the flexible plate with length greater than $2D$ at flexibility $S^{*}=18$, its vibration trajectory course is flared, which is manifested by the extension of the node of second structural mode to both sides of the centerline of the wake flow. \citet{Furquan_Mittal_2021} pointed out that, in this stage of vibration, the second structural mode dominates the oscillation of the splitter plate when the flexibility coefficient is small, while the proportion of the first structural mode increases with the increase of the flexibility. Although the structural modes are orthogonal in theory, the periodic stiffness variations induced by structural deformation, along with the periodic added mass effects from the surrounding fluid, act as modulation mechanisms on the structural response, enabling interaction and coordination among different modes. In particular, for flexible plates, the second natural frequency is approximately six times that of the first natural frequency ($f^{*}_{2}\approxeq 6 f^{*}_{1}$), while the observed resonance frequency locks onto half of the second natural frequency ($f^{*}_{2}/2 \approxeq 3 f^{*}_{1}$). This further supports the existence of such modal coupling effects.

\begin{figure}
	\centerline{\includegraphics[width=13.5cm]{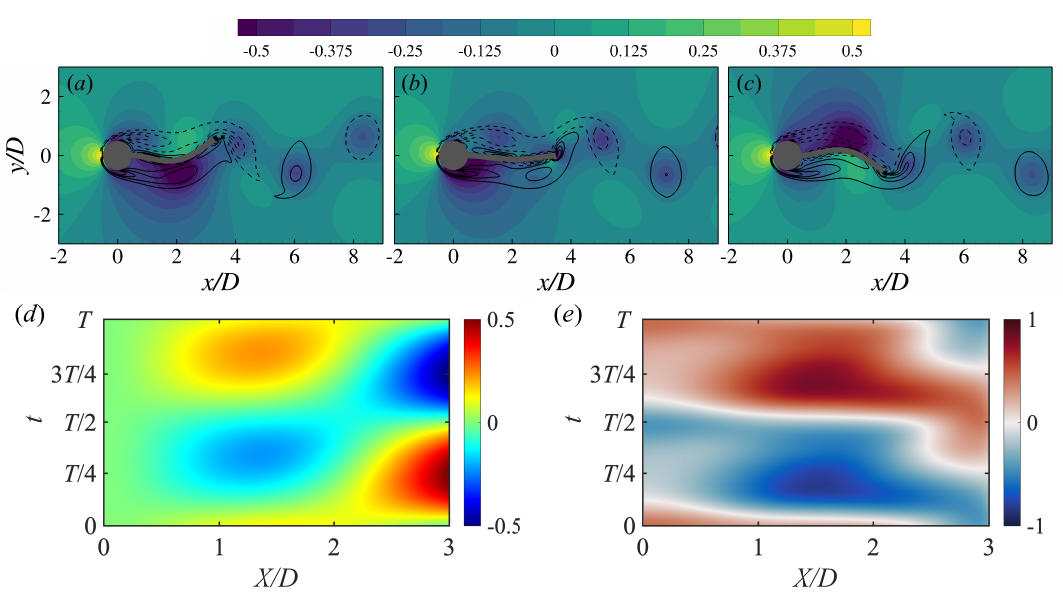}}
	\caption{The displacement variation of the flexible plate and the corresponding pressure field evolve over one vibration cycle for the case at the onset of the symmetry-II stage: contour plots of flow field pressure coefficients snapshots at three moments: when the displacement of the flexible plate tip is at ($a$) its maximum,  ($b$) mean, ($c$) minimum values; and ($d$) transverse displacement $d_{y}(X, t)$, ($e$) pressure difference $p_{d}(X, t)$  variations along plate length $[0, L]$  through oscillation cycle $[\tau, \tau + T]$ for the splitter plate with $L/D=3$ at $S^{*}=7.5$ and $m^{*}=10$.}
	\label{fig20}
\end{figure}

Figure \ref{fig20} illustrates the displacement of the flexible plate and the corresponding flow field variations over one oscillation cycle for the case of flexible plate at $S^{*}=7.5$, $L/D=3$ and $m^{*}=10$.
The selected cycle begins at the moment when the displacement of the flexible plate tip reaches the average displacement point and starts moving in the positive $y$-direction. 
Figures \ref{fig20}($a-c$) presents snapshots of the flow field pressure at three moments: when the displacement of the flexible plate tip is at its maximum, mean, and minimum values, respectively. The corresponding vortex shedding patterns are also depicted as contour lines. 
From the contours, it can be observed that when the plate tip displacement reaches its maximum value, a vortex detaches from the lower surface at the middle of the flexible plate, while the upper surface is positioned between two vortices. Since vortices contain regions of higher fluid kinetic energy and lower pressure, this creates a pressure difference between the upper and lower surfaces of the flexible plate, driving its motion.
Figures \ref{fig20}($d$) and ($e$) show contours of transverse displacement $d_{y}$ and pressure difference $p_{d}$ with $(X, t) \in [0, L]  \times [\tau, \tau + T]$, for the splitter plate with $S^{*}=7.5$ and $L/D=3$.
Here, pressure difference $p_{d}$ can be calculated by $p_{d}=C_{p,lower}-C_{p,upper}$, where $C_{p,lower}$ and $C_{p,upper}$ are pressure coefficient on the upper and lower surface of the splitter plate, respectively.
This periodic vortex shedding generates corresponding periodic pressure difference $p_{d}$ variations along the midspan of the flexible plate, as quantified in figure \ref{fig20}($e$). 
From figures \ref{fig20}($b$) and ($c$), it can be observed that when the flexible plate tip moves downward, the flow velocity in the region beneath the plate is reduced, resulting in a positive pressure in this area. Meanwhile, the vortex above the flexible plate exhibits a more pronounced negative pressure than it did when it was initially shed. This increased negative pressure accelerates the vortex shedding frequency.
By observing figures \ref{fig20}($d$) and ($e$), it is evident that the motion of the middle section of the flexible plate is driven by the pressure difference between the upper and lower surfaces. Additionally, examining the motion of the plate tip reveals that its movement is opposite to the direction of the applied force.
\begin{figure}
	\centerline{\includegraphics[width=6.5cm]{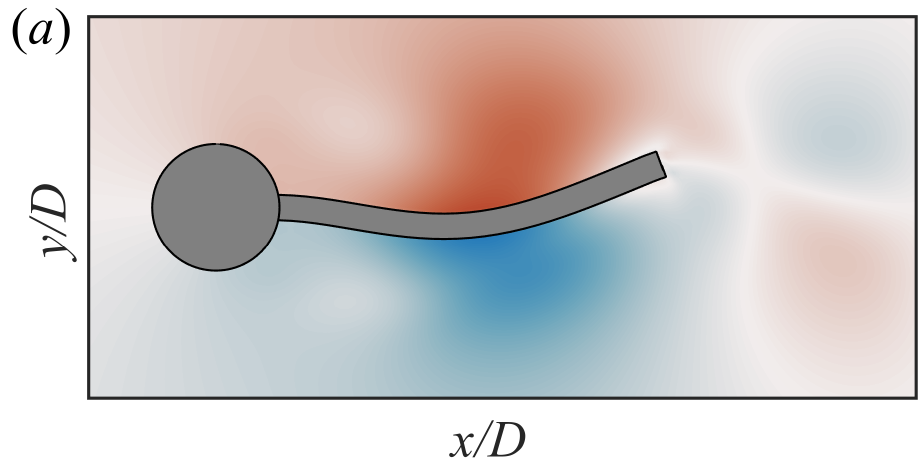}
	           \includegraphics[width=6.5cm]{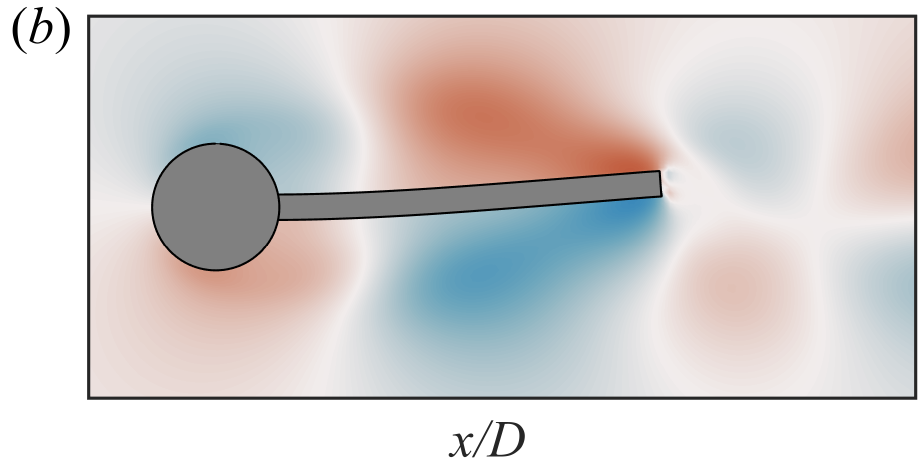}}
	\centerline{\includegraphics[width=10cm]{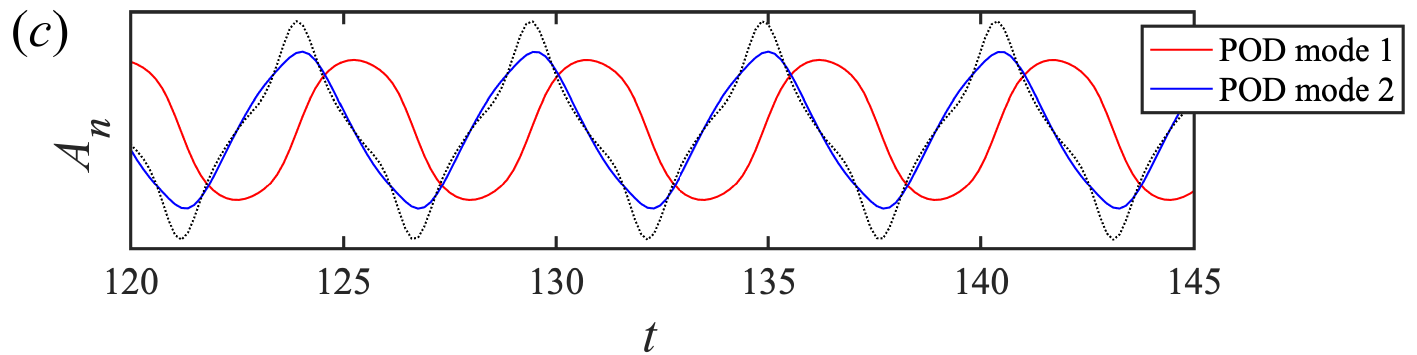}
	}
	
	\caption{An ALE-based POD analysis for the fluctuating pressure field of the moving splitter plate with $L/D=3$ at $S^{*}=7.5$ and $m^{*}=10$: (\textit{a}) POD mode 1 and (\textit{b}) POD mode 2, (\textit{c}) time histories of time correlation coefficients \( A_n \) for the first two POD modes; The dashed line in Figure (\textit{c}) represents the time derivative of the time correlation coefficient of mode 1.  }
	\label{fig241}
\end{figure}

To further investigate the coupling between structural vibration modes, a proper orthogonal decomposition (POD) analysis based on the arbitrary Lagrangian–Eulerian (ALE) mapping was conducted. The POD technique is based on the concept of extracting structures that exhibit the highest correlation with all the velocity fields under consideration \citep{ping2020wake,song2024wake}. In the current work, we first construct a matrix $D$ consisting of $N$ sequential snapshots of the instantaneous fields,
\begin{equation} \label{eq12}
	\mathbf{D}=[\mathbf{d}^{1}~\mathbf{d}^{2}\cdots \mathbf{d}^{N}],
\end{equation}
where $\mathbf{d}^{n}$ represents the $n$-th flow field fluctuations and mesh deformation defined on the orthogonal grid. In each case, a total of $3000$ instantaneous displacement snapshots with a time interval of $\Delta t = 0.1$ were used. Note that $N$ is large enough to ensure the POD results to be well-converged. The resulting autocovariance matrix can be obtained by multiplying the transpose of the flow field snapshot matrix $\mathbf{D}^{T}$ with the snapshot matrix $\mathbf{D}$ itself, as $\mathbf{A}=\frac{1}{N}\mathbf{D}^{T}\mathbf{D}$. Then, by solving an eigenvalue problem $\mathbf{A}V^{i}=\lambda_{i}V^{i}$, we can obtain the eigenvalues $\lambda_{i}$. The POD modes $\phi^{i}$ corresponding to the eigenvalues $\lambda_{i}$ with $\lambda_{1}>\lambda_{2}> \cdots >\lambda_{N}$ can be calculated using the following equation:
\begin{equation} \label{eq13}
	\phi^{i}=\frac{\sum_{n = 1}^{N} V_{n}^{i} \mathbf{d}^{n} }{\left\lVert \sum_{n = 1}^{N} V_{n}^{i} \mathbf{d}^{n} \right\rVert }, i=1,2,\ldots ,N.   
\end{equation}

Figure \ref{fig241} presents the first two POD modes of the fluctuating pressure field for the same case as shown in figure \ref{fig20}. As shown in the figure, the first two POD modes correspond to the second and first structural natural modes, respectively, indicating that the second structural mode dominates under the current parameter conditions. From the fluctuating pressure distribution of the first POD mode, it can be seen that the second structural natural mode is primarily associated with pressure fluctuations in the central region of the flexible plate. This is consistent with the observations in figure \ref{fig20}, where alternating vortex shedding induces periodic pressure differences near the midspan of the plate. Another noteworthy observation is that, in the first POD mode, the direction of the pressure difference across the upper and lower surfaces of the flexible plate aligns with the direction of its motion from the equilibrium position toward the point of maximum displacement. This behavior resembles that of an inverted pendulum model, representing an unstable mode. The temporal coefficients of the two modes exhibit identical frequencies with a phase difference of approximately $\pi/2$, indicating a coupled modal interaction. Such frequency locking and phase relationship persist throughout the symmetric II stage, indicating a coupled behavior between the two modes. The $\pi/2$ phase difference implies that the time derivative of one modal coefficient is either in phase or out of phase with the other mode amplitude variation, thereby ensuring efficient energy transfer between them.

\begin{figure}
	\centerline{\includegraphics[width=7.5cm]{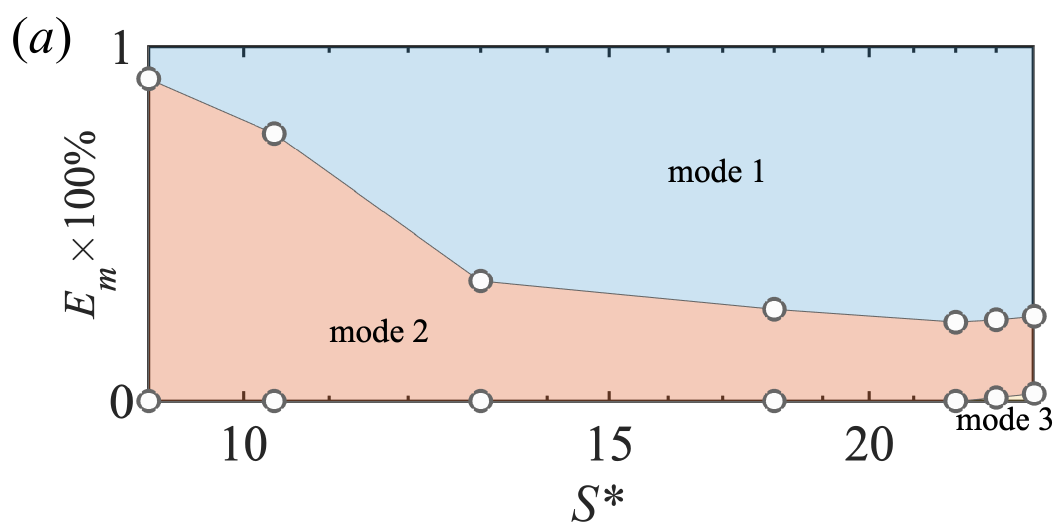}}
	\centerline{\includegraphics[width=4.5cm]{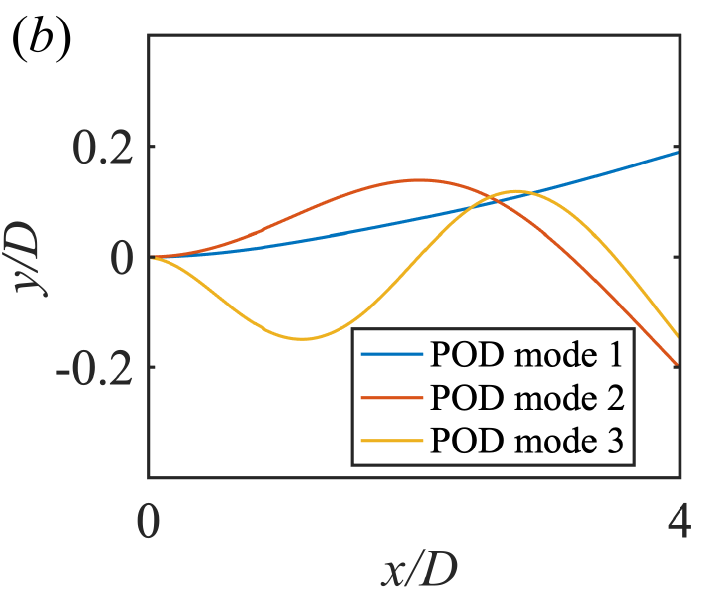}
		\includegraphics[width=4.5cm]{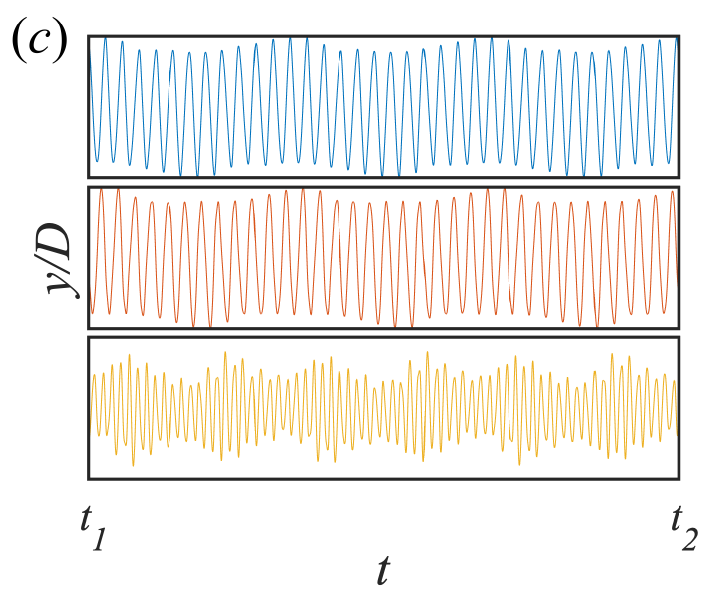}
		\includegraphics[width=4.5cm]{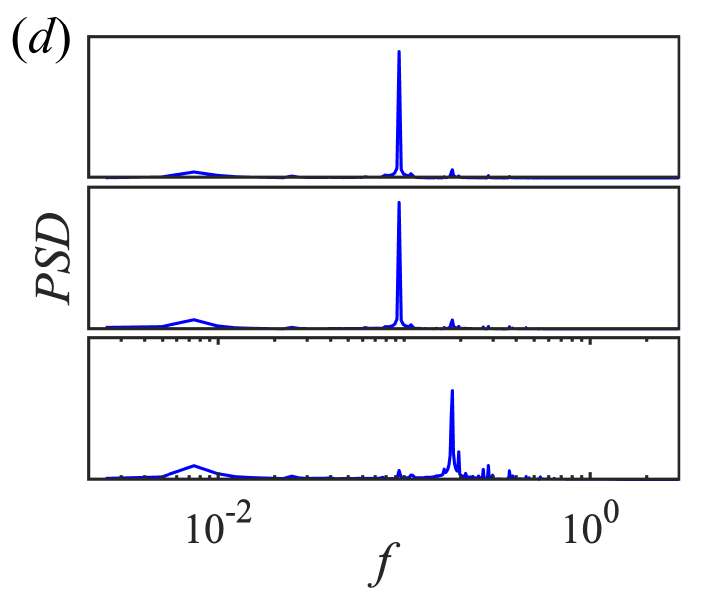}
	}
	
	\caption{A POD analysis for the $L=4D$ cases with increasing flexibility coefficient: (\textit{a}) The kinetic energy contained in the different modes to the total kinetic energy. Results of the POD analysis in case of the flexible plate with $L/D=4$ and $S^{*}=24$: (\textit{b}) transverse displacement distribution along the plate length,   (\textit{c}) time histories of the time correlation coefficients \( A_n \), and (\textit{d}) the frequency distributions of corresponding \( A_n \) for the first three POD modes  }
	\label{fig24}
\end{figure}

\subsection{Quasi-periodic vibration characteristics}
For the cases of the longer flexible plate, a quasi-periodic vibration is observed. To further investigate the effect of increased flexibility on the vibration mode of the flexible plate, we further analyze the differences on dynamic response of the flexible splitter plate with $L=4D$ at different flexibility coefficient $S^{*}$ and $m^{*}=10$ through the proper orthogonal decomposition method. By examining the orthogonal decomposition modes of the structural vibration, several modes resembling the structural vibration shapes of different orders have been identified. Since the eigenvalues $\lambda_{i}$ are proportional to the contribution of the $i$-th mode to the total energy of the system. The kinetic energy contained in the different modes to the total kinetic energy can be calculated used similar POD modes $\phi_{i}$. 
Figures \ref{fig24}($a$) illustrates the energy distribution among the structural modes in the oscillations of flexible splitter plate as a function of flexibility coefficient $S^{*}$.
From the figure, it can be observed that when the flexibility coefficient $S^{*}=9$, the vibration of the structure is dominated by the second structural mode, which is consistent with previous research \citep{Furquan_Mittal_2021}. Furthermore, as the flexibility coefficient increases, the contribution of the second structural mode gradually diminishes, while the first structural mode increasingly dominates the oscillations of the flexible splitter plate. When the flexibility coefficient exceeds 22, the structural vibrations exhibit the presence of a minor third structural mode. The emergence range of the third structural mode precisely corresponds to the symmetry-III stage. This indicates that the third structural mode may be responsible for inducing quasi-periodic vibrations in the structure. Figures \ref{fig24}($b$)-($d$) present the vertical displacement distributions of the first three POD modes, the time correlation coefficients $A_{n}$, and the frequency distribution of $A_{n}$ for the flexible plate case with the plate length of $L=4D$ and flexibility coefficient of $S^{*}=24$. It can be observed in figure \ref{fig24}($d$) that the first and second POD modes of the structure still exhibit same frequencies, whereas the dominant frequency of the third POD mode is 1.9 times the first two vibration frequency. Another notable observation is that the temporal coefficients of the first two POD modes also maintain a phase difference of $\pi/2 $. Moreover, the third POD mode exhibits substantial low-frequency components, thereby confirming its role in inducing low-frequency vibrations within the structure.

\begin{figure}
	\centerline{\includegraphics[width=13.5cm]{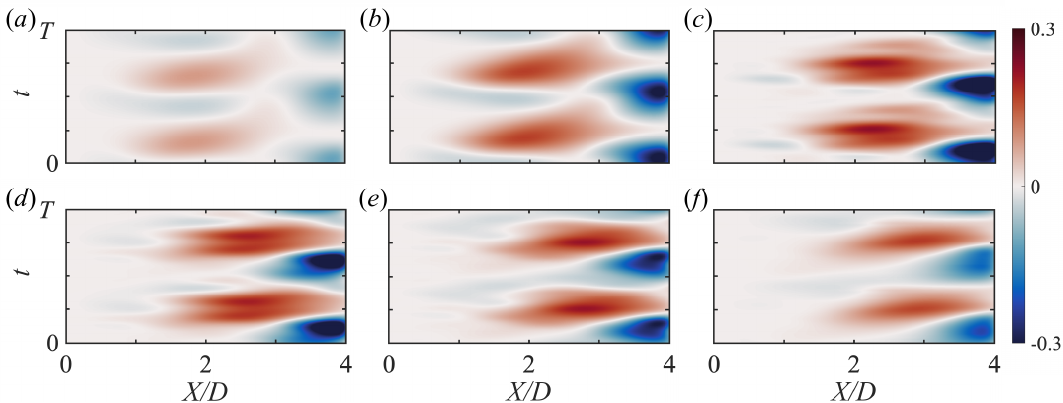}}
	\caption{
		Contour plots of the energy transfer rate $P(X, t)$ variations along plate length $[0, L]$  through oscillation cycle $[\tau, \tau + T]$ for the splitter plate with $L/D=4$ at increasing flexibility coefficient ($a$) $S^{*}=9$,  ($b$) $S^{*}=10.35$, ($c$) $S^{*}=13$, ($d$) $S^{*}=18$, ($e$) $S^{*}=22$ and ($f$) $S^{*}=24$.}
	\label{fig26}
\end{figure}

The difference from the case at the onset of the symmetric-II stage is explored by the power of the flow field doing work on the flexible plate. Figure \ref{fig26} shows contours of energy transfer rate $P(X, t)$ with $(X, t) \in [0, L]  \times [\tau, \tau + T]$, for the splitter plate with $L/D=4$ at different $S^{*}$. 
Energy transfer rate (per unit length), $P(X, t)$, at a location $X$ along the splitter plate centerline and
instant $t$ is defined as follows:
\begin{equation} \label{eq7}
	P(X, t)= (\textbf{u}_{s}(X, t) \cdot \vec{n})  p_{d}(X, t)
\end{equation}
where ${d}(X, t)$ and $\vec{n}$ respectively denote the velocity vector and the normal vector at $X$ in the local coordinate of the flexible splitter. The viscous forces contribute negligibly to the energy budget and are not included in the estimate.  Because the case with a flexibility coefficient of 24 exhibits quasi-periodic vibration characteristics, the variations in pressure differences were obtained through phase averaging method.
As previously described, for cases with lower $S^{*}$ in the flexible plate, the structural vibration is primarily dominated by the second structural mode. 
It primarily absorbs energy from the flow field through displacements occurring in the midspan of the flexible plate, while the motion at the trailing end dissipates energy. 
For the more flexible cases, a noticeable change is that the area where the flexible plate absorbs energy moves towards the tail of the flexible plate. 
It can be observed that a distinct difference in cases dominated by the first-order mode is the presence of an energy-absorbing region at the tail, where energy is extracted from the flow field. With increasing flexibility, the amplitude of structural vibration grows, accompanied by enhanced energy exchange efficiency. However, upon entering the quasi-periodic regime, despite a further increase in vibration amplitude, the efficiency of energy transfer decreases. This may represent an undesirable condition for certain energy harvesting devices.


\section {\label{sec4}Flow field characteristics in the symmetry-breaking stage }
\subsection {Mean flow field characteristics}
\begin{figure}
	\centerline{\includegraphics[width=13.5cm]{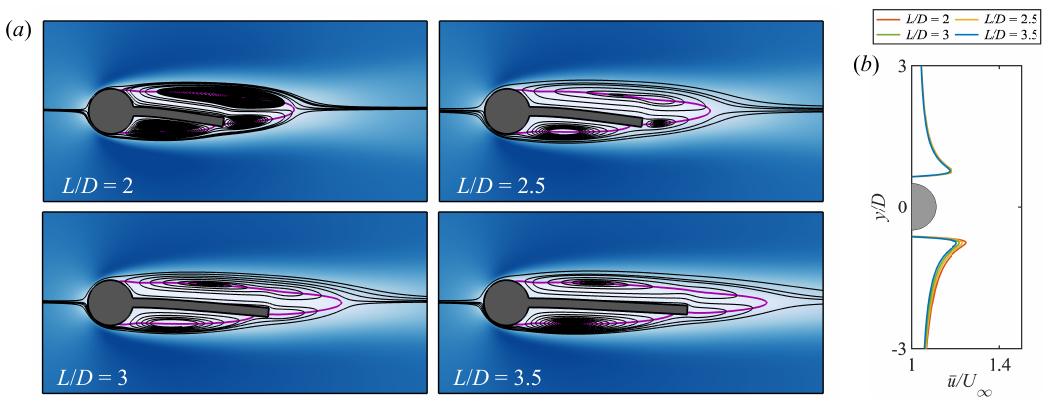}}
	\caption{Time-averaged velocity distribution for cases in bifurcation-I stage: ($a$) contour plots of the mean streamwise velocity with the time-averaged streamlines, ($b$) mean streamwise velocity profiles at $x/D=0$ for the cases with different plate length at the largest flexibility coefficient cases in the bifurcation-I stage. In sub-figure ($a$), the contour lines where the streamwise velocity is equal to zero are shown in magenta color. The mean streamwise velocity $\overline{u}$ is limited in a range from $-0.1$ (white) to $1$ (blue).}
	\label{fig10}
\end{figure}

Previous studies \citep{shukla2009flow,wu2014flow,Furquan_Mittal_2021} have detected the symmetry-breaking phenomenon in flow-structure interaction in cylinder wake flow. 
In the present numerical simulations, the symmetry-breaking phenomenon is observed in cases where the plate length is less than $4D$. When the flexible plate deflects to one side, the flow field is inevitably altered. To investigate this effect, we first examine the distribution of the time-averaged streamwise velocity during the bifurcation-I stage.
Figure \ref{fig10}($a$) presents the contour plots of the mean streamwise velocity for the largest flexibility coefficient cases with different plate length in the bifurcation-I stage, where the time-averaged vertical displacement of the plate tip nearly reaches its maximum value.
In addition, the streamlines adjacent to the solid boundary and a magenta line representing the zero streamwise velocity contour are also depicted in the figures.
For the sake of facilitating comparisons of flow field statistics across different plate length cases, the deflection direction of the flexible plates are mirrored toward the negative $y$-direction in this section.
The wake recirculation region is observed to extend with increasing plate length.
At $L = 2D$, the wake exhibits an asymmetric structure: a comparatively larger recirculation bubble on the positive $y$-side, and on the negative $y$-side, two smaller bubbles formed by the deformed plate—one just behind the cylinder and another near the trailing edge.
As the splitter plate length increases, the recirculation bubble on the positive $y$-side gradually divides into two distinct bubbles. The upstream bubble enlarges and shifts further upstream. 
Because the separated shear layers are attracted by the negative pressure within the recirculation zone, the width of the recirculation region in the far wake gradually decreases. The splitter plate tip lies mainly along the boundary of the recirculation region, where the local velocity is nearly zero.
Previous results of \citet{sahu2023symmetry} indicate that, after filtering out fluctuations in the flow field, the splitter plate gradually shifts toward the boundary of the recirculation zone as flexibility increases. This can explain why the deflection positions of longer separation plates undergoing symmetry-breaking phenomena are closer to the wake centerline. From figure \ref{fig10}($b$), it can be observed that there are marked velocity differences between the two sides of the cylinder.
Such a velocity difference generates a pressure differential between the two sides of the flexible plate, thereby maintaining its deflected position. At the same time, the streamline plots show that, on the high-velocity side, the flow separates near the tip of the flexible plate and, after passing through the tip vortex, merges into the recirculation zone on the opposite side. This process partially compensates for the flow deficit on that side. As the length of the flexible plate increases, the strength of the tip vortex gradually decreases, and the velocity difference between the two sides of the plate is correspondingly reduced.

\begin{figure}
	\centerline{\includegraphics[width=13.1cm]{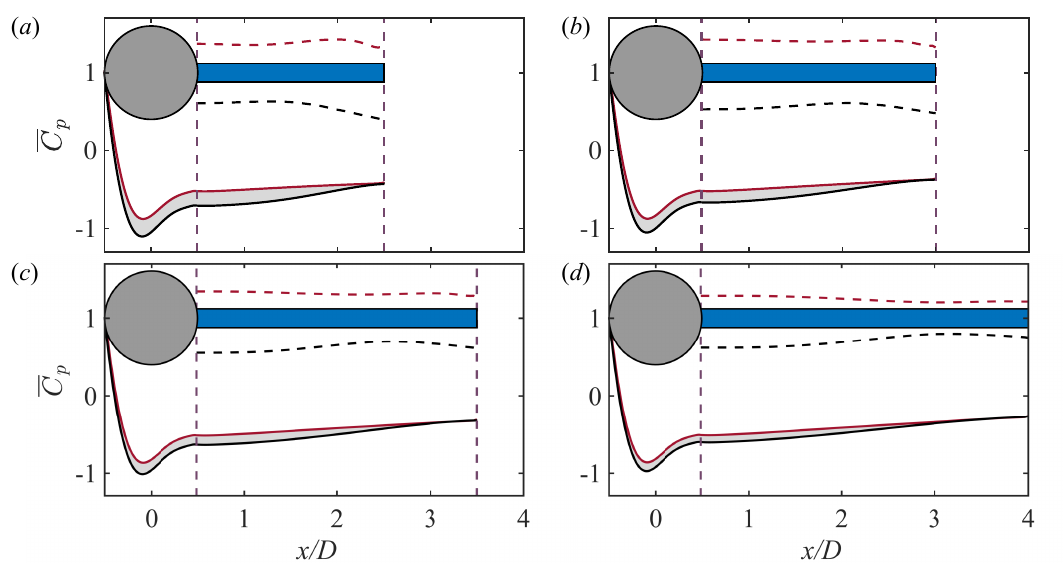}}
	\caption{Time-averaged pressure distributions on the solid surface for cases in the bifurcation-I stage with ($a$) $L/D=2$,  ($b$) $L/D=2.5$,  ($c$) $L/D=3$,  ($d$) $L/D=3.5$ . The RMS pressure values  (magnify tenfold) on both sides of the flexible plate are illustrated with dashed lines. The red line represents the upper surface of the solid structure, the black line represents the lower surface, and the pressure difference is filled in gray. }
	\label{fig12}
\end{figure}

Figure \ref{fig12} shows the time-averaged pressure distributions on the solid surface for the symmetry-breaking cases correspond to those shown in Figure \ref{fig10}($a$). The instantaneous pressure coefficient $C_{p}$ is defined as $C_{p}= (p - P_{\infty})/(0.5\rho U_{\infty}^{2})$, where $p$ is the local static pressure on the solid surface and $P_{\infty}$ is the free stream pressure. Additionally, the RMS pressure values on both sides of the flexible plate are illustrated with dashed lines.
On the side toward which the flexible plate deflects, larger negative pressures are observed. The pressure difference between the two sides supports the internal restorative stresses generated by the deformation of the flexible plate. However, as the plate length increases, this pressure difference gradually diminishes.
Unlike the mean pressure coefficient, which maintains a similar distribution shape across different plate lengths, the RMS pressure values in cases with shorter plates exhibit a distinct peak at the plate tip on the deflection side. As the plate length increases, this tip peak gradually vanishes.
This may be because shorter plates are more strongly influenced by the shear layers separated from the circular cylinder surface. In contrast, for longer plates, even though the plate tip is closer to the boundary of the recirculation bubble, it is farther from the shear layer separation point. As a result, the local velocity gradient is reduced, and the flow field is more stable.
Another piece of evidence indicating that the flow field becomes more stable as the plate length increases is the reduction in pressure fluctuations near the main cylinder observed in the $L/D=3.5$ case.

\subsection {Linear stability analysis}

\begin{figure}
	
	\centerline{
		\includegraphics[width=6.5cm]{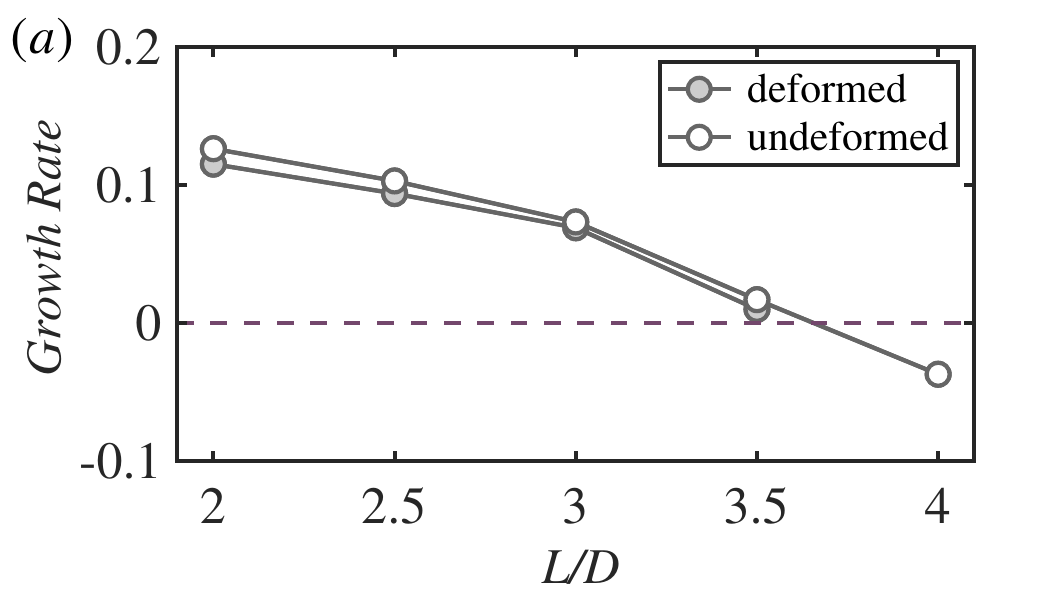}
		
		\includegraphics[width=6.5cm]{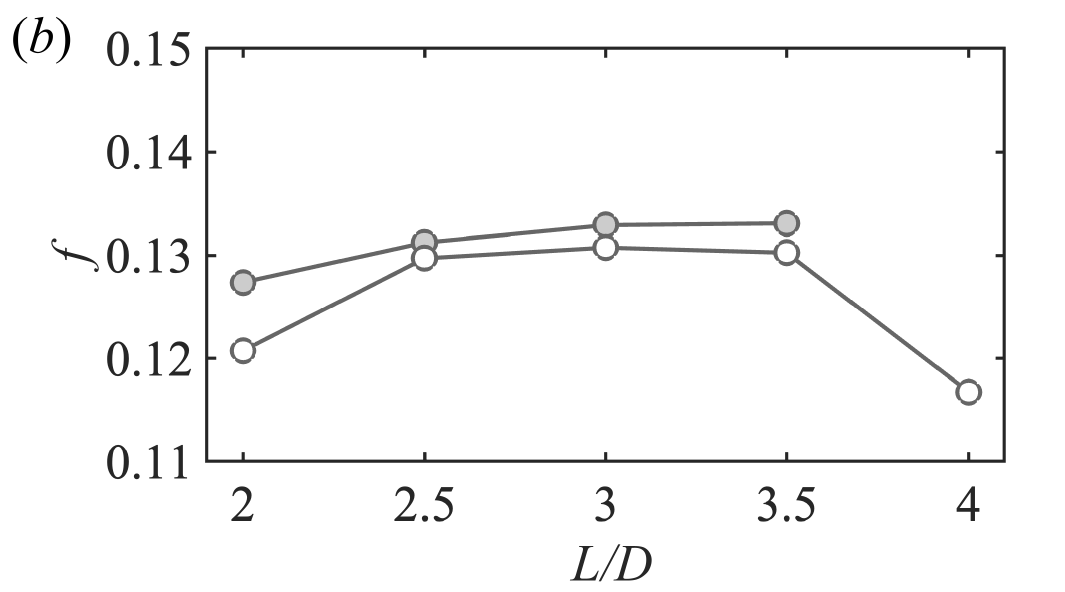}
	}
	
	\caption{($a$) Growth rates and ($b$) frequencies of dominant unstable modes of the cylinder with an attached plate with different plate length.}
	\label{fig13}
\end{figure}

To validate the effect of the splitter plate on flow stability, a linear stability analysis is performed on the flow field of an undeformable splitter plate attached to the rear of a cylinder in laminar flow.
The analysis considers both undeformed splitter plates of varying lengths fixed at the wake centerline and those fixed at their mean deflected positions in bifurcation-I stage.
Figure \ref{fig13} shows the growth rates and frequencies of the dominant instability modes in the flow field for different cases.
The deflected cases correspond to those shown in Figure \ref{fig10}, with the same flexibility coefficient.
The higher the growth rate of the instability mode, the more unstable the flow field becomes. When the growth rate exceeds zero, the flow field becomes unstable, leading to alternating vortex shedding in the wake.
When the plate length is $4D$, the growth rate is less than zero, indicating that the flow field is in a completely symmetric steady state about the wake centerline. This explains the phenomenon where stiff flexible plates also exhibit steady-state flow fields.
Even when deflection occurs and the flow symmetry is broken, the instability does not increase. On the contrary, the growth rate is lower than in the case where a splitter plate of the same length is fixed at the wake centerline. This further supports the physical rationale of the symmetry-breaking phenomenon.
For cases where the growth rate is greater than zero, the frequency of the unstable modes increases with the length of the splitter plate. However, for cases where the unstable modes decay, the frequency decreases to approximately 0.116. This decaying mode has been confirmed to play a crucial role in inducing instability in the wake of the flexible splitter plate.

\subsection{Re-entry into symmetry-breaking bifurcation stage}

\begin{figure}
	
	\centerline{
		\includegraphics[height=4cm]{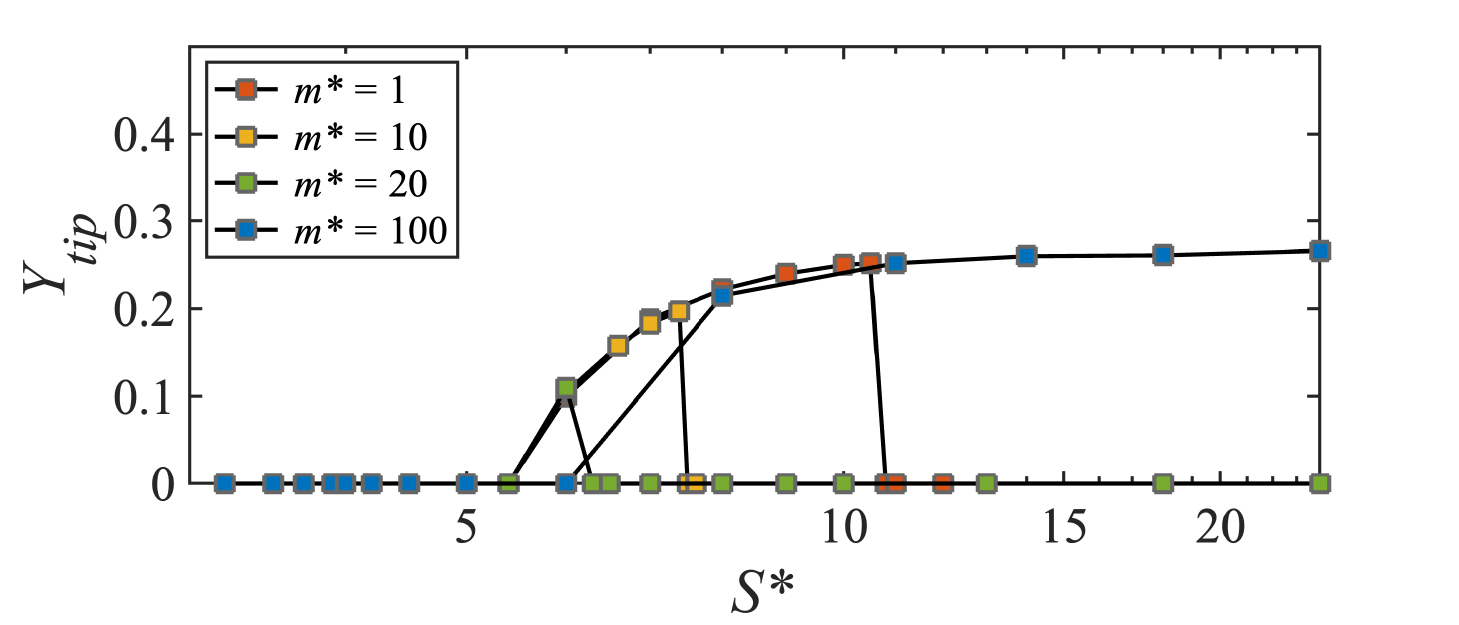}
	}
	
	\caption{  Symmetry breaking deflection in term of the time-averaged position $Y_{tip}$ at the plate tip, with increasing flexibility coefficient $S^{*}$ for the cases with different $m^{*}$ and $L/D=3$.}
	\label{figmy}
\end{figure}

For the cases with the shorter flexible plate of $L=2D$ and $m^{*}=10$, an interesting phenomenon is the reappearance of the symmetry breaking bifurcation as the flexibility coefficient continues to increase after experiencing the symmetry-II stage of fluid-solid interaction vibrations. The research of \cite{nc} shows that for a freely rotating cylinder attached with a rigid plate, the maximum lateral rotation angle caused by symmetry breaking is only related to the length of the plate. For the flexible plate cases, the structural stiffness acts as an additional restoring force. Ideally, a flexible plate with a length less than $4D$ should have a gradually increasing lateral displacement as the flexibility increases until it reaches a maximum value determined only by the flow field characteristics. Figure \ref{figmy} shows the variation of the time-averaged lateral displacement at the tip of the $3D$-long flexible plate with respect to the flexibility coefficient under different mass ratios $m^{*}$. As shown in the figure, the evolution of $Y_{\text{tip}}$ for flexible plates with different mass ratios follows a consistent trend in the regime where fluid–structure resonance does not occur. This indicates that the development of symmetry breaking in the flexible plate depends only on the characteristics of the flow field and the structural stiffness. For the flexible plate with $L/D=3$ and $m^{*}=100$, resonance causes the vibration frequency to initially lock onto a sub-harmonic of the second natural frequency. As the flexibility increases further, the frequency re-locks onto the base vortex frequency, marking the end of the resonance.

\begin{figure}
	
	\centerline{
		\includegraphics[width=10cm]{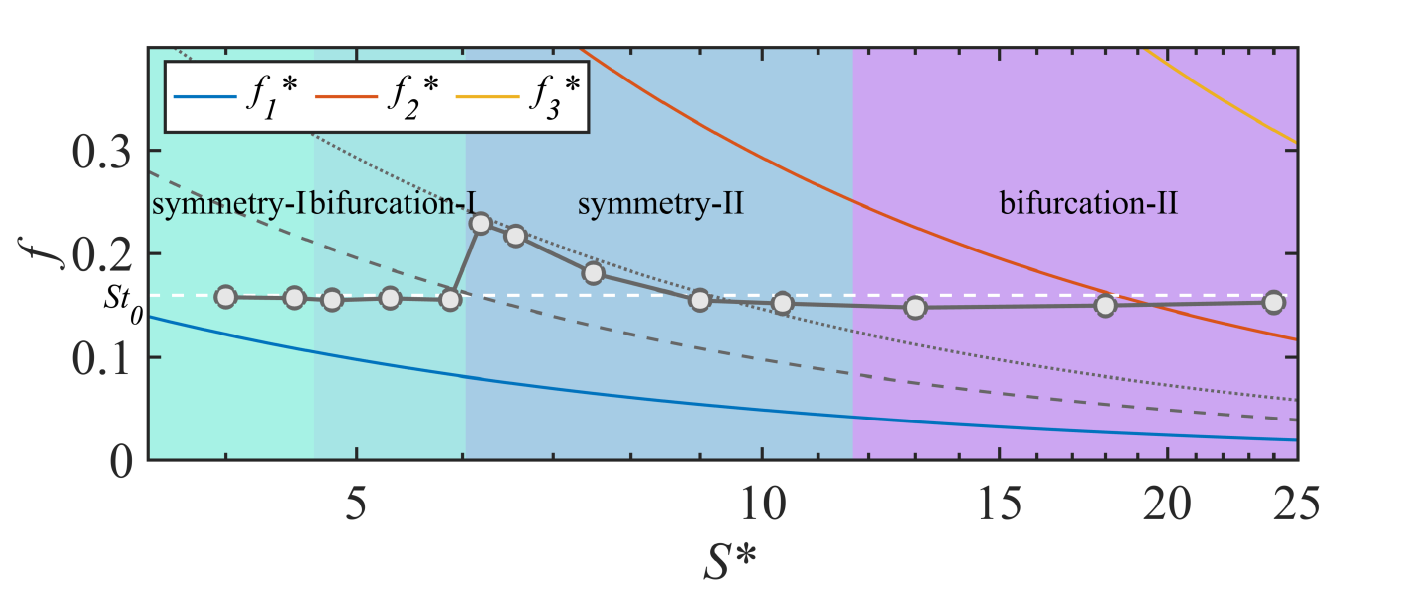}
	}	
	\caption{Variation of dominant frequency and natural frequencies $f^{*}_{i}$ of the flexible splitter plate with $L/D=2$ as a function of flexibility coefficient $S^{*}$. The grey dashed line represents twice the first natural frequency, and the grey dotted line is a sub-harmonic (half value) of the second-order natural frequency. The base vortex frequency $St_{0}$ obtained from the rigid plate case is indicated by the white dashed line, Different vibration mode stages are distinguished by different background colors.}
	\label{fig18}
\end{figure}

Figure \ref{fig18} shows the variation of the tip vibration frequency of the flexible plate with respect to the flexibility coefficient $S^{*}$ for the cases  with $L/D=2$ and $m^{*}=10$. Similar to the cases with larger mass ratios, the case with a plate length of $2D$ and a mass ratio of 10 also exhibits a re-locking of the vibration frequency to the base vortex frequency, although without a distinct frequency jump. This phenomenon is accompanied by the reappearance of symmetry breaking. In fact, for the case with flexibility coefficient equal to $10.35$, which is the case where the plate vibration frequency starts to deviate from the sub-harmonic of the second natural frequency, the maximum amplitude of the structure starts to decrease. These observations confirm that shorter flexible plates, similar to cases with higher mass ratios, tend to shorten the resonance interval. This behavior may be explained by two factors: first, shorter plates are less efficient at extracting energy from vortex shedding; and second, they are more susceptible to thickness-related effects, which enhance internal damping.

\section{\label{sec6}Conclusion}


This study investigates the complex fluid–structure interaction (FSI) dynamics of a flexible splitter plate attached to the rear of a circular cylinder at a low Reynolds number of $Re=150$. The effects of the plate length $L/D$ and flexibility coefficient $S^{*}$ and mass ratio $m^{*}$ on the wake dynamics, plate vibrations, and flow stability are systematically explored. 

The wake dynamics and oscillations of the flexible plate vary significantly with $L/D$ and $S^{*}$. Linear stability analysis reveals that, as the rigid plate length increases, the growth rate of the dominant instability mode gradually decreases. When the splitter plate length reaches $4D$, the growth rate becomes negative, indicating a steady flow without vortex shedding.
Based on the equilibrium position and tip amplitude, five distinct response modes of the flexible plate are identified. The symmetry-I stage appears for $L/D \leq 3.5$ at low bending stiffness, whereas a steady flow with a stationary splitter plate precedes the symmetry-II stage for $L/D = 4.0$. During symmetry-I, vortex shedding from the cylinder boundary layer induces small-amplitude vibrations of the splitter plate in its first structural mode. For higher bending stiffness, large-amplitude vibrations occur in the symmetry-II and symmetry-III stages, accompanied by excitation of the second structural mode.
A symmetry-breaking bifurcation-I stage occurs between the symmetry-I and symmetry-II modes. For shorter plates ($L/D \leq 2.5$), a unique re-entry into the bifurcation-II stage is observed as $S^{*}$ increases further after transitioning through symmetric vibration stages. In the bifurcation stages, plate deflection induces asymmetry in the wake and significantly elongates the recirculation region.
The 2S vortex-shedding pattern is identified in most cases, whereas the P+S mode is observed only for short plates ($L = 2D$) with low bending stiffness during the symmetry-II stage. As vibration amplitude increases, the main cylinder and the plate tip generate two large vortices, each carrying two smaller vortices in the wake, resembling the 2P mode.


The mechanism underlying the formation of large-amplitude, strongly coupled fluid–structure excitation has been elucidated.
Proper orthogonal decomposition (POD) analysis based on the arbitrary Lagrangian–Eulerian mapping reveals that, during resonance, the first and second structural modes remain synchronized at the same frequency with a fixed phase difference of $\pi/2$. Furthermore, by examining the variation of vibration frequency with the flexibility coefficient across different mass ratios—while excluding added mass effects—frequency locking to the subharmonic of the second natural frequency ($f^{*}_{2}/2$) is confirmed.
These findings indicate that resonance arises from the coupling between the first and second structural modes, driven by the combined effects of fluid-added mass and periodic stiffness variations. For flexible plates with high mass ratios or for longer plates at large $S^{*}$, the third structural mode is more likely to be excited. Its excitation is accompanied by a P+S wake pattern oscillating vertically at low frequencies, resulting in quasi-periodic structural vibrations during the symmetry-III stage.
A reduction in plate length and an increase in mass ratio are both associated with a narrower resonant locking range. Consequently, symmetry-breaking phenomena can be observed in cases with high flexibility coefficients, even though greater flexibility generally promotes more intense vibrations. Finally, the symmetry-breaking phenomenon is found to depend primarily on plate length and stiffness, suggesting that, in the absence of resonance, lateral displacement gradually approaches its maximum value as flexibility increases.

The present study contributes to the knowledge of intricate interactions between flexible structures and wake flows and its practical applications in engineering and biomimetic design. Understanding frequency lock-in and vibration patterns can optimize the design of flexible structures for extracting energy from fluid flows. Longer flexible plates can stabilize wake flows, offering a potential strategy for controlling fluid-induced vibrations in engineering systems. The current linear instability analysis does not consider the vibration of the flexible plate, and future linear stability analysis that consider the dual physical fields of fluids and solids will help to understand the transition from steady state fields to large amplitude vibrations.

\section{Acknowledgments}
The research was funded by the National Key R\&D Program of China (No. 2023YFE0120000), the National Natural Science Foundation of China (No. 52478535, 12402208), Guangdong Basic and Applied Basic Research Foundation (No. 2023A1515240054), Program for Intergovernmental International S\&T Cooperation Projects of Shanghai Municipality, China (No. 22160710200), the Natural Science Foundation of Chongqing, China (CSTB2023NSCQ-MSX0060).
\section{Declaration of interests} 
The authors report no conflict of interest.
\section{Data avaliblity}
The data that support the findings of this study are available from the corresponding author upon reasonable request.

\bibliography{jfm}

\bibliographystyle{jfm}


\end{document}